\documentclass[twocolumn]{aastex62}

\newcommand{\zreion}{z_{\mathrm{re}}}
\newcommand{\HI}{H{\sc~i}}
\newcommand{\HII}{H{\sc~ii}}
\newcommand{\HeI}{He{\sc~i}}
\newcommand{\HeII}{He{\sc~ii}}

\newcommand{\Treion}{T_{\rm reion}}
\newcommand{\vIF}{v_{\mathrm{IF}}}
\newcommand{\spec}{\alpha_{\mathrm{IF}}}
\newcommand{\kpc}{\mathrm{kpc}}
\newcommand{\Mpc}{\mathrm{Mpc}}

\usepackage{color}

\shorttitle{Heating of the IGM by Reionization}
\shortauthors{D'Aloisio et al.}

\begin{document}

\title{Heating of the Intergalactic Medium by Hydrogen Reionization}

\author{Anson D'Aloisio}
\email{ansond@ucr.edu}
\affil{Department of Physics \& Astronomy, University of California, Riverside, CA 92521, USA}

\author{Matthew McQuinn}
\affil{Astronomy Department, University of Washington, Seattle, WA 98195, USA}

\author{Oliver Maupin}
\affiliation{Department of Physics and Astronomy, Haverford College, 370 Lancaster Avenue, Haverford, PA 19041, USA}

\author{Frederick B. Davies}
\affiliation{Department of Physics, University of California, Santa Barbara, CA 93106-9530, USA}

\author{Hy Trac}
\affiliation{McWilliams Center for Cosmology, Department of Physics, Carnegie Mellon University, 5000 Forbes Avenue, Pittsburgh, PA 15213, USA}

\author{Spencer Fuller}
\affiliation{Department of Physics, University of California, Davis, One Shields Ave. Davis, CA 95616, USA}

\author{Phoebe R. Upton Sanderbeck}
\affiliation{Astronomy Department, University of Washington, Seattle, WA 98195, USA}

\begin{abstract}
During reionization, the intergalactic medium is heated impulsively by supersonic ionization fronts (I-fronts).  The peak gas temperatures behind the I-fronts, $\Treion$, are a key uncertainty in models of the thermal history after reionization.  Here we use high-resolution radiative transfer simulations to study the parameter space of $\Treion$.  We show that $\Treion$ is only mildly sensitive to the spectrum of incident radiation over most of the parameter space, with temperatures set primarily by I-front speeds.  We also explore what current models of reionization predict for $\Treion$ by measuring I-front speeds in cosmological radiative transfer simulations.  We find that the post-I-front temperatures evolve toward hotter values as reionization progresses.  Temperatures of $\Treion = 17,000-22,000$ K are typical during the first half of reionization, but $\Treion = 25,000 - 30,000$ K may be achieved near the end of this process if I-front speeds reach $\sim10^4~\mathrm{km/s}$ as found in our simulations.  Shorter reionization epochs lead to hotter $\Treion$.  We discuss implications for $z>5$ Ly$\alpha$ forest observations, which potentially include sight lines through hot, recently reionized patches of the Universe.  Interpolation tables from our parameter space study are made publicly available, along with a simple fit for the dependence of $\Treion$ on the I-front speed.        
\end{abstract}

\keywords{intergalactic medium -- dark ages, reionization, first stars -- cosmology: theory}

\section{Introduction}

Nearly all of the hydrogen in the Universe was reionized and heated by the rise of the first galaxies and quasars.  Cosmic Microwave Background (CMB) measurements place the midpoint of this process at $z\approx 8-9$ \citep{2016A&amp;A...596A.108P}, while Ly$\alpha$ forest observations show that it must have been mostly complete by $z=6$ \citep{mcgreer15}.  When combined with other probes such as the visibility of Ly$\alpha$ emitting galaxies and quasar damping wing analyses, these observations suggest that the end of reionization was likely near $z=6$ \citep[e.g.][]{2007MNRAS.381...75M,2010ApJ...723..869O,2010MNRAS.408.1628S,2011ApJ...743..132P,2013MNRAS.429.1695B,2014MNRAS.443.2831C,2014MNRAS.437.2542T,2014arXiv1412.4790C,2015MNRAS.446..566M, 2018ApJ...856....2M, 2011Natur.474..616M,2018Natur.553..473B,2018arXiv180206066D}.  Further constraining the timing and duration of reionization would provide insight into the first sources of ionizing radiation in the Universe (see \citealt{2016ARA&A..54..313M} for a recent review).   

One path towards constraining reionization is to look for its heating effects on the intergalactic medium (IGM).  As ionization fronts (I-fronts) expand supersonically around the first ionizing sources, they impulsively heat the gas to temperatures between $15,000-30,000$ K \citep{1994MNRAS.266..343M, 1997MNRAS.292...27H, 2004MNRAS.348..753S, 2007MNRAS.380.1369T, 2008ApJ...689L..81T, 2011MNRAS.417.2264V, 2012MNRAS.426.1349M, 2018arXiv180500099F}.  We shall refer to the temperature achieved by this impulsive heating as the {\it post-I-front temperature}, denoted by $\Treion$.  After the I-front passes, the gas is driven to cooler temperatures over cosmological timescales primarily by the expansion of the Universe and inverse Compton scattering with CMB photons.  

The impact of reionization on the thermal history of the IGM is in principle detectable in the small-scale structure of the Ly$\alpha$ forest \citep[e.g.][]{2010ApJ...718..199L, 2011MNRAS.410.1096B, 2012MNRAS.424.1723G, 2017PhRvD..96b3522I}. The volume-weighted mean temperature rises steadily during reionization, reaches a peak near the end of this process, and declines before the onset of \HeII\ reionization \citep[e.g.][]{2016MNRAS.460.1885U, 2017arXiv171204464D, 2018arXiv180104931P}. The detection of a rise in temperature with redshift at $z>5$ would be a tell-tale sign of reionization, and would place constraints on its timing and the nature of its sources.  It may also be possible to detect spatial variations in the temperature owing to the inhomogeneity of reionization \citep{2008ApJ...689L..81T,2009ApJ...706L.164C,furlanetto09,2014ApJ...788..175L, 2015ApJ...813L..38D, 2018MNRAS.tmp..938K}.  However, theoretical predictions for both of these signatures are highly uncertain.  
  
\begin{figure}
\centerline{
\includegraphics[width=8.5cm]{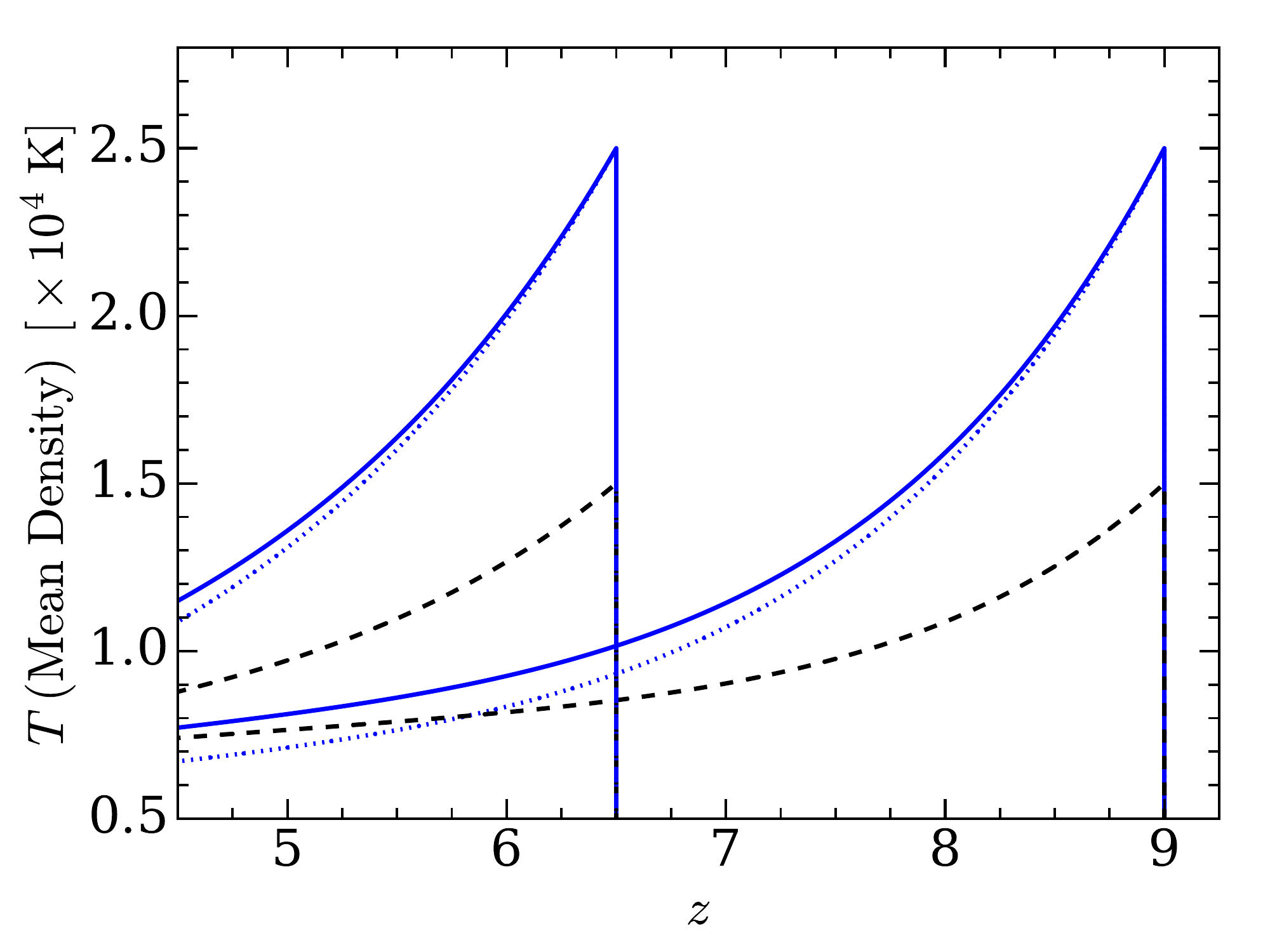}
}
\caption{The impact of uncertainties in $\Treion$ on IGM temperatures at $z\gtrsim5$.  We model the impulsive heating from a passing I-front by instantaneously heating gas parcels to a temperature $\Treion$.  Here we show two illustrative sets of examples in which mean-density parcels are reionized at $z=6.5$ and $9$.  We vary $\Treion$ from 15,000 (dashed) to 25,000 K (solid), representative of the range of values that have been assumed in the literature.  Uncertainties in $\Treion$ translate to large uncertainties in the expected temperature of recently reionized gas.  For comparison, another significant source of uncertainty in the gas temperature is the post-reionization photoheating rate, which is set by the spectral index of the ionizing background, $\alpha$. We have assumed $\alpha=1$, but the dotted curves show the modest effects of assuming $\alpha = 2$ for the case with $\Treion=25,000$ K.}
\label{FIG:T_intro}
\end{figure}

Post-I-front temperatures are the chief source of uncertainty for theoretical models of the thermal history.   To illustrate this point, Fig. \ref{FIG:T_intro} shows the thermal histories of mean-density gas parcels that are impulsively heated to different temperatures.  The dashed and solid curves assume $\Treion = 15,000$ and $25,000$ K, respectively, representative of the range of values that are found in the literature \citep[e.g.][]{2012MNRAS.426.1349M, 2018arXiv180104931P, 2018arXiv180500099F}. For the post-reionization photoheating rate, we assume a power-law spectrum such that $J_{\nu} \propto \nu^{-\alpha}$ with $\alpha = 1$ ($J_{\nu}$ is the specific intensity and $\nu$ is frequency).\footnote{In detail, we assume that the spectrum has a sharp cutoff at 4 Ry, which is motivated by models in which stellar emissions dominate the ionizing background at the redshifts of interest \citep[e.g.][]{2009ApJ...703.1416F, 2012ApJ...746..125H, 2018arXiv180104931P}.}  Though the effects of $\Treion$ are modest for the gas parcels reionized at redshift $\zreion=9$, there are $50$ \% differences in the $z= 5.5$ temperatures for those reionized at $\zreion=6.5$.  Note also that larger $\Treion$ leads to larger temperature {\it dispersion}, which can be seen from the spread between parcels reionized at $z=6.5$ and 9.    For comparison, the dotted curves adopt $\alpha=2$ for $\Treion=25,000$ K, illustrating the effects of uncertainties in the {\it post-reionization} photoheating rate.\footnote{In \S \ref{sec:spec}, we will show that $\alpha=1-2$ is within the range of expectation from theoretical models of young stellar populations.}  Variations of $\Delta \alpha = 1$ lead to modest $\lesssim 15$ \% differences in temperature at $z\sim 5.5$.  This simple illustration suggests that $\Treion$ plays an important role in interpreting $z>5$ Ly$\alpha$ forest measurements, which may probe recently reionized patches of the IGM. 

In principle, it is possible to mitigate these uncertainties because the physics that determines $\Treion$ is well known \citep{1994MNRAS.266..343M}.  The characteristic width of I-fronts during reionization is several times the mean free path through the neutral gas.  The spectrum of incident radiation sets the maximum possible temperature, $3 k_{\mathrm{B}} T \sim 13.6~\mathrm{eV}/(\alpha -1)$, e.g. $T\sim 50,000$ K for $\alpha = 2$. However, collisionally excited (\HI\ Lyman-series) line cooling by neutrals within the I-front can cool the gas to significantly lower temperatures, depending on how long it spends inside the front.  This, in turn, depends on how quickly the I-front is moving, which is set by the number flux of ionizing photons on the front boundary.  (As we will show, the actual temperatures are rarely higher than $30,000$ K.)  In summary, $\Treion$ depends mainly on the spectrum of the ionizing radiation and the speeds at which the fronts are moving.  

The above discussion highlights why $\Treion$ remains so uncertain.  First, without observational constraints on the sources and sinks of ionizing photons during reionization, neither the spectrum of the background nor the I-front speeds are known with certainty.  Secondly, given the short spatial and time scales associated with the I-fronts, it is not obvious that cosmological radiative transfer (RT) simulations are converged with respect to $\Treion$.  In this paper, we present a focused study that aims to improve our understanding of the $\Treion$ parameter space.  This study will inform future Ly$\alpha$ forest measurements by making clearer the connection between IGM temperatures and the nature of the reionization process.  

Following \citet{1994MNRAS.266..343M}, a number of authors have explored $\Treion$ \citep{2007MNRAS.380.1369T, 2008ApJ...689L..81T, 2011MNRAS.417.2264V, 2012MNRAS.426.1349M, 2018arXiv180500099F}.   Some of these studies have reached different conclusions about the likely values of $\Treion$. For example, \citet{2012MNRAS.426.1349M} used 1D RT simulations to argue that post-I-front temperatures should be in the range $\Treion=20,000-30,000$ K. On the other hand, \citet{2018arXiv180500099F} examined $\Treion$ values in the Technicolor Dawn cosmological RT simulations and found $\Treion = 14,000-19,000$ K. In addition to the lack of consensus amongst past studies, the exact dependence of $\Treion$ on the incident spectrum and the I-front speed has not been explored in detail. In this paper, we expand upon previous works by performing the first systematic study of the $\Treion$ parameter space.  For this task, we use a suite of high-resolution, 1D RT simulations to ensure that our results are numerically converged in $\Treion$. After defining the parameter space, we then apply stellar population synthesis modeling, and a set of cosmological RT simulations, to explore what contemporary models of reionization predict for $\Treion$.    

The remainder of this paper is organized as follows.  In \S \ref{sec:post-I-front}, we present our parameter space study of $\Treion$.  In \S \ref{sec:model_expectations}, we attempt to narrow this parameter space using expectations from current models of ionizing source spectra and I-front speeds during reionization.  In \S \ref{sec:implications}, we discuss the implications of our results for the thermal history of the IGM and for high-$z$ Ly$\alpha$ forest measurements.   We offer concluding remarks in \S \ref{sec:conclusion}.  Unless otherwise noted, all distances and velocities are quoted in {\it physical} units.

\begin{figure*}
\centerline{
\includegraphics[width=14cm]{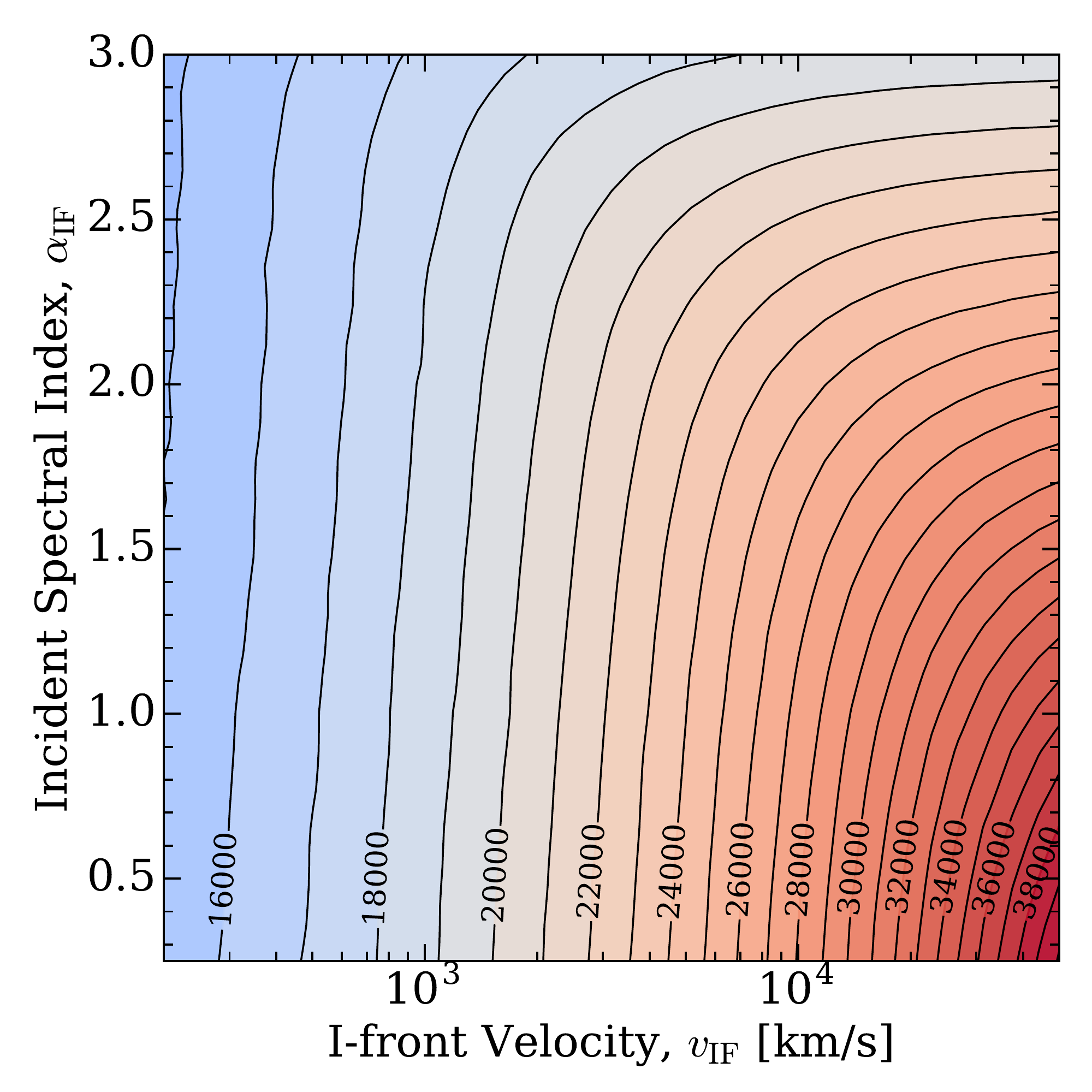}
}
\caption{The parameter space of post-I-front temperatures during reionization.  The curves show contours of constant $\Treion$ in Kelvin.  The $y$-axis corresponds to the spectral index of the incident ionizing radiation, while the $x$-axis corresponds to the proper I-front speed.  The contours are insensitive to redshift and gas density for the bulk of the low-density IGM (see top panel of Fig. \ref{fig:Treion_density}).  In \S \ref{sec:model_expectations}, we use stellar population synthesis modeling to argue that spectral indices of $\spec \lesssim 1.5$ are most relevant during reionization. Using cosmological RT simulations of reionization, we find that the I-fronts are slow during the early stages of reionization, with speeds ranging from $10^2 - 2\times10^3$ km/s, but near the end of reionization they reach speeds $\sim 10^4$ km/s.     }
\label{fig:Treion_contour}
\end{figure*}

\section{Post-I-front Temperatures}
\label{sec:post-I-front}

In this section, we present our parameter space study of $\Treion$. We begin by describing the numerical methodology of our calculations.   

\subsection{Numerical Methodology}
\label{sec:methodology}

Our calculations are based on the 1D RT code of \citet{2016MNRAS.457.3006D}, which employs the numerical approach of \citet{2007MNRAS.374..493B}.  We refer the reader to those papers for technical details.  In summary, the code tracks the propagation of ionizing radiation from a point source into hydrogen and helium gas at a fixed redshift.  The 1D RT, ionization balance, and temperature equations are solved on a uniform spatial grid with cell size $\Delta x = 1$ proper kpc.  We initialize the gas to a temperature of 100 K, but our results are insensitive to this choice.  The gas density is uniform, set to the cosmic mean of the epoch under consideration.  { In Appendix \ref{sec:density_flucs}, we present test runs with skewers taken from cosmological simulations. The results from those runs indicate that our main conclusions would be unchanged in the presence of density fluctuations.  We will elaborate upon this important point in the next section.}  

The spectrum of ionizing radiation is discretized over 25 logarithmically spaced frequency bins between 1 and 4 Ry.  In Appendix \ref{sec:convergence} we demonstrate the numerical convergence of our results with respect to the spatial cell size and frequency binning.  We adopt a power-law spectrum characterized by the spectral index $\spec$, such that the specific intensity of the radiation is $J_{\nu} \propto \nu^{-\spec}$, where $\nu$ is frequency.  The sharp cutoff at energies above 4 Ry is motivated by the standard assumption that stellar sources dominated the ionizing photon budget during reionization \citep[e.g.][]{1987ApJ...321L.107S,2009ApJ...703.1416F,2012ApJ...746..125H,2013MNRAS.436.1023B,2017MNRAS.468.4691D}.  We neglect secondary ionizations, which have an insignificant effect for the adopted source spectrum.  The code includes \HI, \HeI, and \HeII\ photoheating, and all of the relevant cooling processes for intergalactic gas of primordial composition: collisional excitation, adiabatic expansion, Compton, recombination, free-free, and collisional ionization.  The code assumes that the timescale for photoelectrons to thermalize with the neutrals in an I-front is much shorter than the time that the gas spends inside the front. (In fact, to our knowledge, all previous calculations have adopted this assumption.)  We justify this assumption in Appendix \ref{sec:thermalization}.    

We have run a suite of RT simulations spanning a range of source luminosities and $\spec$.  We measure I-front speeds ($\vIF$) directly from the simulations by tracking the location of the $x_{\mathrm{HI}} = 0.5$ boundary with time.  In a given run, the I-front begins at its fastest speed and slows down with time, since the ionizing flux scales as $r^{-2}$ (where $r$ is the distance from the point source). This allows us to sample a range of $\vIF$ within a single simulation, and our runs vary the source luminosities to achieve a wider range of $\vIF$.  

As noted above, our primary goal is to quantify how $\Treion$ depends on $\vIF$ and the spectral index of the ionizing radiation.  Two things complicate the use of the aforementioned RT code for this purpose: (1) At a given I-front location, we would like to extract the gas temperature immediately after the I-front has passed to avoid the onset of cooling processes.  However, as we consider I-fronts of various speeds and widths, there is no single, robust prescription for when $\Treion$ should be measured; (2) Absorption by residual neutral gas in equilibrium between the source and the I-front tends to harden the spectrum of  radiation impinging on the I-front.   The degree of hardening will vary with distance to the source, and with its luminosity. To make the interpretation of $\Treion$ unambiguous, we would like to roll these effects into the parameter $\spec$, such that it is the spectral index of the radiation that is {\it incident} on the I-front.  

We have developed a method of circumventing both of these issues at once.  We modified the RT code to turn off Hubble and Compton cooling, as these processes dominate the cooling {\it after} the gas leaves the I-front, operating over cosmological time-scales.\footnote{The characteristic time that gas spends inside an I-front is given by equation (\ref{EQ:tIF}). For $\vIF = 1,000~(10,000)$ km/s (velocities spanning most of reionization; see \S \ref{sec:velocities}), $t_{\mathrm{IF}}\sim$ 10 (1) Myr. }  In addition, we turn off all thermal evolution for post-I-front gas that has reached \HI\ ionization equilibrium, i.e. cells for which $(\Delta n_e/n_\mathrm{tot}) (\Delta x/c/dt)< 10^{-8}$, where $\Delta n_e$ is the change in the electron number density in time step $dt$, $n_\mathrm{tot}$ is the number density of gas particles (atoms, ions, electrons), and $c$ is the speed of light.  Importantly, this preserves the relevant heating and cooling processes while the gas is still inside the I-front, but after the I-front passes the temperature remains fixed, allowing us to simply measure $\Treion$ from the last simulation output.   To reduce the effects of spectral hardening, we also set the neutral fraction of this equilibrium gas to an arbitrarily low value.  In this limit, the radiation that is incident on the I-front has the same spectral index as that of the source, $\spec$. From here on we will identify $\spec$ with the incident radiation.  We have tested our $\Treion$ results against those from the original version of the code, verifying that they are consistent in the regime where spectral hardening is negligible for the latter (see Appendix \ref{sec:convergence}). 

{ Lastly, we note that our 1D RT code adopts an infinite speed of light for computational efficiency.  Previous studies have shown that this approach provides an exact solution to the I-front propagation as observed along the line of sight to the source, in the case with finite speed of light \citep{2003AJ....126....1W,2006ApJ...648..922S, 2007MNRAS.374..493B, 2016MNRAS.457.3006D}.  We can therefore apply the same transformation between apparent and actual I-front velocities to translate our infinite-speed-of-light $\vIF$ to the case with finite speed of light.  We obtain the true I-front speeds using the relation $\vIF = v_{\mathrm{IF},c=\infty}/(1+v_{\mathrm{IF},c=\infty}/c)$, where $v_{\mathrm{IF},c=\infty}$ is the speed measured in the simulation \citep{2006ApJ...648..922S}.  In practice, this correction only comes into play for I-front speeds near the end of reionization.}

\subsection{Results}  
\label{sec:Treion}

\begin{figure}
\centerline{
\includegraphics[width=8.7cm]{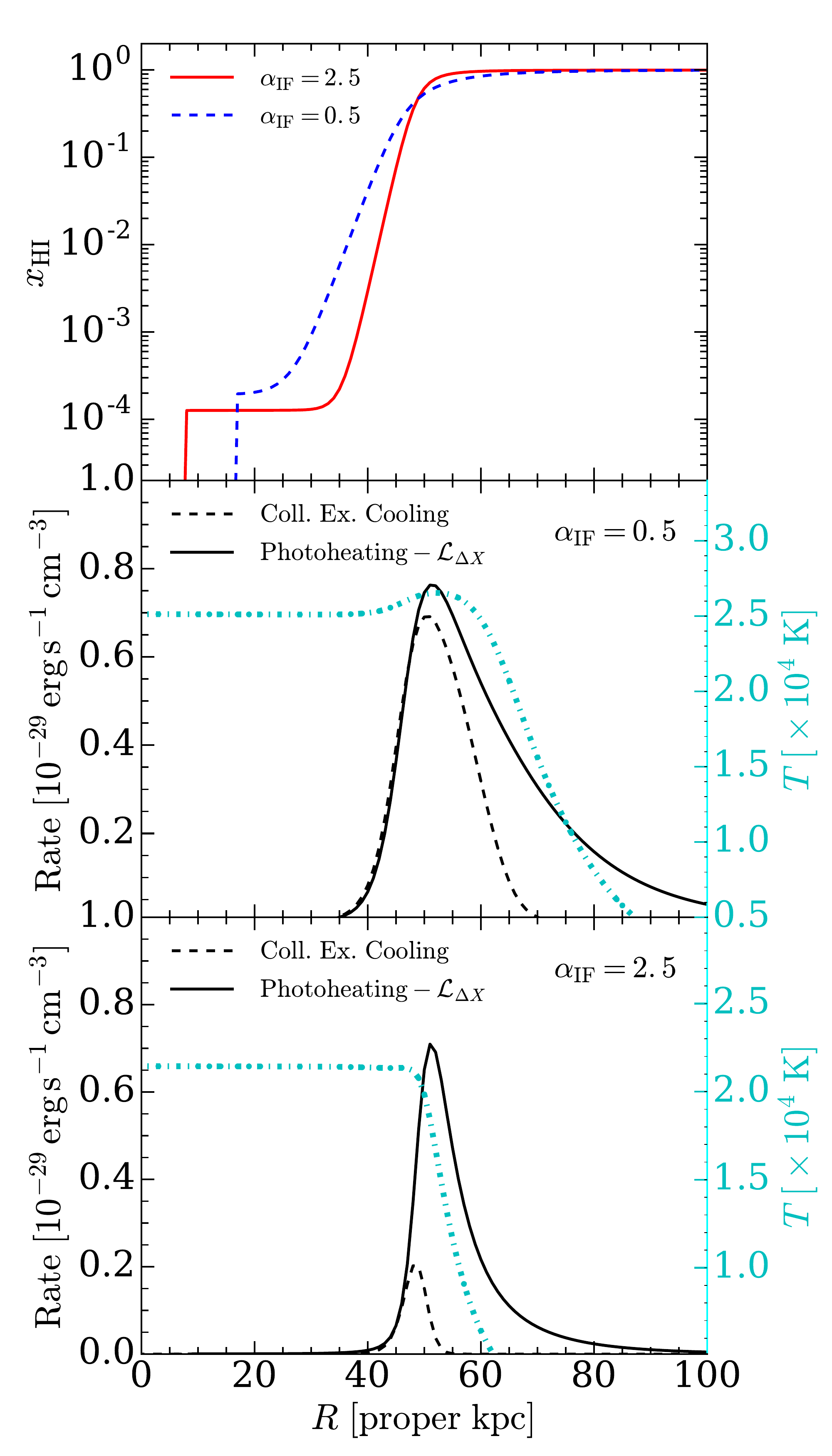}
}
\caption{Ionization and thermal structures of I-fronts.  Top panel: the \HI\ fraction as a function of distance for two cases illustrating that I-fronts are broader for harder spectra.   The solid and dashed lines correspond to $\spec = 2.5$ and $0.5$, respectively.    The $x$-axes have been shifted such that $x_{\mathrm{HI}}=0.5$ at $R=50$ kpc, and both fronts are traveling at $\vIF \approx 6,000$ km/s at $z=6$.  The sharp cutoffs in $x_{\mathrm{HI}}$ owe to our methodology for rolling the effects of spectral hardening into the parameter $\spec$ (see \S \ref{sec:methodology}).  Middle and bottom panels:  the corresponding temperatures (cyan dot-dashed/right-axis), \HI\ $+$ \HeI\ photoheating rates (solid) and  collisional excitation cooling rates (dashed).  (See text for discussion on $\mathcal{L}_{\Delta X}$.) The photoheating from a hard incident spectrum is compensated by the increased collisional excitation cooling rate (owing to its steep dependence on temperature) over a wider area. }
\label{fig:I-front_structure}
\end{figure}

Figure \ref{fig:Treion_contour} shows the main result of this paper: the post-I-front temperature ($\Treion$) as a function of the spectral index of incident radiation ($\spec$) and I-front speed ($\vIF$).  The curves correspond to contours of constant $\Treion$ at $z=6$ and $\Delta = 1$, where $\Delta$ is the gas density in units of the cosmic mean.  A distinguishing feature of Fig. \ref{fig:Treion_contour} is that $\Treion$ is only mildly sensitive to $\spec$ over much of the parameter space, especially in the hard spectrum limit (small $\spec$).  This behavior is the result of an interplay between the photoheating and the cooling that occurs in the boundary layer of the I-front.   Collisional line excitation cooling is exponentially sensitive to temperature and is most efficient when there are equal numbers of neutrals and electrons.  Making the spectrum harder increases the energy injection into the gas and broadens the I-front such that there is more collisional line cooling.  

To illustrate these effects, Fig. \ref{fig:I-front_structure} shows the ionization and thermal structures for I-fronts with $\spec=2.5$ and $0.5$, and $\vIF \approx 6,000$ km/s.  The top panel shows the \HI\ fractions, $x_{\mathrm{HI}}$, where the $x$-axes have been shifted such that $x_{\mathrm{HI}}=0.5$ at $R=50$ kpc.  Note that the sharp cutoffs in the top panel owe to our procedure for rolling the effects of spectral hardening into $\spec$, and the gas temperatures are not allowed to evolve to the left of those cutoffs (as described in \S \ref{sec:methodology}). The bottom and middle panels show the corresponding temperatures, as well as  \HI\ $+$ \HeI\ photoheating and collisional excitation cooling rates.   For the former, we have subtracted off $\mathcal{L}_{\Delta X} = (3/2) k_{\mathrm{B}} dn_{\mathrm{tot}}/dt$ (where $k_\mathrm{B}$ is Bolztmann's constant), which accounts for the fact that the heat must be distributed amongst nearly twice the particles in the newly ionized gas (see eq. {\ref{eq:Tevol}}).  For $\spec=0.5$, there is more photoheating over a broader path length.  In the absence of cooling, this would result in significantly hotter temperatures compared to the case with $\spec=2.5$.  However, the steep dependence of the cooling rate on temperature and the wider I-front conspire to keep the increase in $\Treion$ modest.  

{
The bottom two panels in Fig. \ref{fig:I-front_structure} show that collisional excitation cooling nearly matches the effective heating rate from photoionizations behind the front.  This suggests that the post I-front temperature can be calculated by solving the balance equation for heating and cooling.  Although our RT code solves the full non-equilibrium equations, let us explore the accuracy of the equilibrium assumption by writing
\begin{equation}
C(T) x_{\rm HI} n_e - \frac{3}{2} k_b T  \frac{d x_{\rm HI}}{dt} = \Delta E_{\rm HI} \Gamma x_{\rm HI}. 
\end{equation}
where  $C(T)$ is the collisional cooling rate coefficient, $\Gamma$ is the photoionization rate, and $\Delta E_{\rm HI}$ is the excess energy per photoionization, which sufficiently behind the front should take the optically thin value. Approximating $x_{\rm HI}$ as $\exp[-\Gamma t]$ behind the front, we may write this as
\begin{equation}
C(T) n_e + \frac{3}{2} k_b T ~ \Gamma = \Delta E_{\rm HI} \Gamma. \label{eqn:secondbalance}
\end{equation}
Note that equation (\ref{eqn:secondbalance}) is independent of $x_{\rm HI}$; there is only one equilibrium temperature during this exponential phase. Also, the equation depends on density only through $v_{\rm IF} \propto \Gamma/ n_\mathrm{e}$.     

We find that that if we solve equation~(\ref{eqn:secondbalance}) for the equilibrium temperature, including terms for \HeI\ that were omitted above for brevity, the solution is accurate to a couple thousand Kelvin in the lower left quadrant of Fig. \ref{fig:Treion_contour} (corresponding to where there is sufficient heating from the harder spectrum, and more time owing to the slower front speeds, to establish this equilibrium).  However, equation~(\ref{eqn:secondbalance}) does not work as well for the other parameter space, undershooting $\Treion$ by $5-10,000$ K at high $\vIF$, implying that the detailed heating within the I-front matters there.  That $\Treion$ is in many situations set by equilibrium behind the I-front indicates that the  relevant distance scale for numerically resolving $\Treion$ is larger than a few mean free paths.  In this regime, the relevant scale is the distance over which a few photoionization timescales occur behind the front, or $ v_{\rm IF}  \Gamma^{-1} \approx (\sigma_0 n_H)^{-1} (3 +\alpha)/\alpha$, where $\sigma_0$ is the photoionization cross section of hydrogen at 1 Ry.  This expression is a factor of order ten larger than a naive estimate based on the mean free path, $(\sigma_0 n_H)^{-1}\sim1~$physical kpc.  This may explain why simulations are able to roughly capture $\Treion$ for much of reionization if they have resolutions of $\sim 10$~physical kpc.  Note, however, that this regime becomes less applicable towards the end of reionization, when I-fronts are moving at their fastest speeds. }        

The contours in Fig. \ref{fig:Treion_contour} are insensitive to redshift over the range of interest for reionization.  This property may be understood using a simple scaling argument from \citet{2016MNRAS.457.3006D}, and noting that the main effect of redshift in our homogeneous simulations is to rescale the gas density.  Consider an I-front with instantaneous speed $\vIF$.  As noted above, the gas in the front would be heated to some maximum temperature determined by $\spec$ in the absence of cooling.  However, line cooling will lower the temperature by an amount $\Delta T \sim t_{\mathrm{IF}} L_{\mathrm{cool}}/n_{\mathrm{H}}$, where $t_{\mathrm{IF}}$ is the time spent within the front and $L_{\mathrm{cool}} \sim n_{\mathrm{H}}^2$ is the cooling rate.  For fixed $\vIF$, the time spent in the front is $t_{\mathrm{IF}}\propto R_{\mathrm{IF}}/\vIF \propto 1/(n_{\mathrm{H}} \vIF)$, where $R_{\mathrm{IF}}$ is the front width, implying that $\Delta T \propto 1/\vIF$.  It follows then that $\Treion$ should be insensitive to $z$ {\it at fixed $\vIF$}.

\begin{figure}
\includegraphics[width=8.7cm]{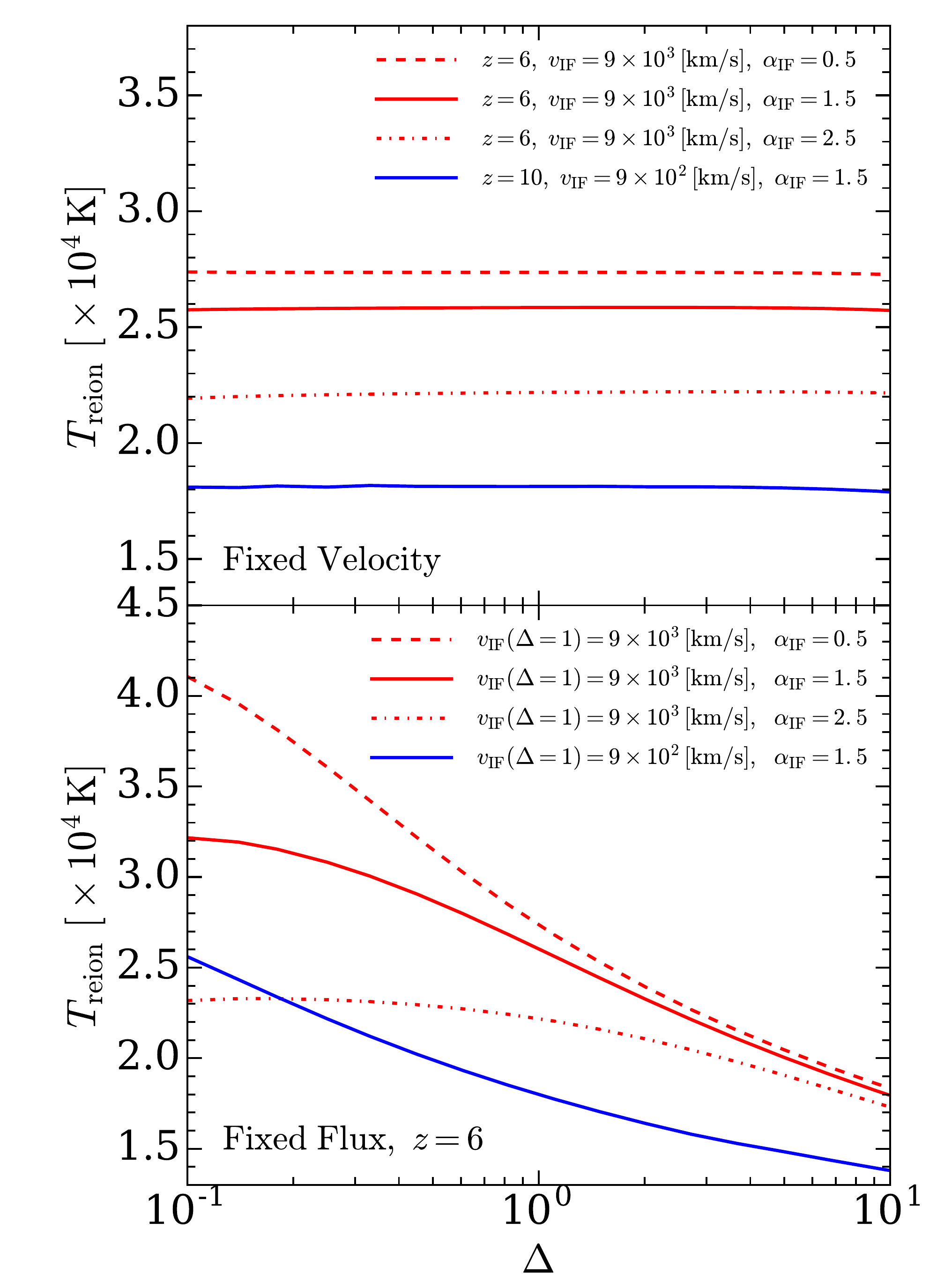}
\caption{Dependence of post-I-front temperatures on gas density.  Top panel: $\Treion$ vs. $\Delta$ for fixed I-front speed.  Post-I-front temperatures are insensitive to the gas density for $\Delta =0.1-10$ {\it at fixed $\vIF$}.  The top three curves illustrate the effect of varying $\spec$ from $\spec = 0.5$ to $2.5$.  The bottom curve considers a slower front speed at $z=10$.  Bottom panel: $\Treion$ vs. $\Delta$ for fixed ionizing flux (i.e. the front speed varies with local density).  The I-fronts accelerate through (slow down in) regions of under(over)-dense gas, resulting in a hotter (cooler) temperatures.  This inverted temperature-density relation is a potential signature of recently reionized gas. }
\label{fig:Treion_density}
\end{figure}

The above argument implies that $\Treion$ should also be insensitive to $\Delta$ as long as the recombination time is much longer than the collisional excitation cooling time.  We have explored this dependence numerically by performing a set of RT simulations in which the source luminosity is varied to keep $\vIF$ fixed over the range of $\Delta=0.1-10$.  The top panel of Fig. \ref{fig:Treion_density} shows the results of these runs.  The dashed, solid, and dot-dashed curves show the dependence of $\Treion$ on $\Delta$ for $\spec = 0.5$, $1.5$, and $2.5$, respectively.  For these runs, we adopt $z=6$ and a fixed I-front speed of $\vIF = 9 \times 10^3$ km s$^{-1}$.  The blue solid curve corresponds to a slower speed of $\vIF=9\times 10^{2}$ km s$^{-1}$ at $z=10$ with $\spec=1.5$. (In \S \ref{sec:velocities}, we will find that these two speeds are representative of their corresponding redshifts.)   The main point is that $\Treion$ depends weakly on $\Delta$ at fixed $\vIF$, consistent with the simple scaling argument of the last paragraph.

Next we examine the dependence of $\Treion$ on the local gas density {\it for a fixed ionizing flux}.  As I-fronts sweep through the IGM, density variations in the cosmic web modulate the local I-front speeds, accelerating through under-densities and slowing within over-densities. Thus, recently reionized gas should exhibit an inverted temperature-density relation with hotter (cooler) temperatures corresponding to under(over)-dense gas.  Note that this is not the same as the inversion that owes to gas parcels being reionized at different times.  In the latter case, over-dense regions tend to be colder because they are reionized earlier, so they have had more time to cool \citep[see e.g.][]{2008ApJ...689L..81T, 2009ApJ...701...94F}.  In contrast, the inversion under consideration here applies to gas that is reionized at nearly the same time, and owes entirely to the density dependence of I-front speeds. To investigate the magnitude of this effect, we have run a series of RT simulations at fixed ionizing flux spanning a range of $\Delta$ at $z=6$ (near the likely end of reionization). As our primary motivation is exploring the signature of recently reionized gas in high-$z$ quasar absorption spectra, we neglect shock heating, which is insignificant at the densities considered \citep{2016MNRAS.456...47M}.  The bottom panel of Fig. \ref{fig:Treion_density} shows the $\Treion$-$\Delta$ relation derived from our simulations.  To connect these results to the contour plot in Fig. \ref{fig:Treion_contour}, we denote these runs by their I-front speeds at $\Delta = 1$.  As anticipated, the post-I-front temperature decreases with density, reflecting the fact that I-fronts travel slower through larger $\Delta$.  For example, $\Treion$ varies by $\approx 5,000$ K between $\Delta = 0.3$ and $3$.  The trend is stronger (weaker) for harder (softer) spectra.     

{ Lastly, in Appendix \ref{sec:density_flucs}, we have also explored the impact of density fluctuations using sight lines extracted from a high-resolution cosmological simulation.  The tests presented there indicate that the contours in Fig. \ref{fig:Treion_contour} would be unchanged in the presence of density fluctuations.  This lack of sensitivity results from the fact that the relevant heating and cooling processes at a given location within the I-front depend only on the optical depth of the gas behind the location; they are independent of the structure of the intervening gas.   } 

Based on the results of this section, we are led to conclude that $\Treion$ can be determined for most of the intergalactic gas if the local $\vIF$ and, to a lesser extent, $\spec$ are specified.  Indeed, following the results in Fig. \ref{fig:Treion_contour}, $\Treion$ may be determined to within $1000$ K for $\spec \lesssim 1.5$ using the I-front speed alone.  We provide a five-parameter polynomial fit to $\ln(\Treion)$ at fixed $\spec =1.5$, 

\begin{equation}
\ln(\Treion) = \sum_{n=0}^4 C_n \ln\left(\frac{\vIF}{\mathrm{[km/s]}}\right)^n,
\label{eq:Treion_fit}
\end{equation} 
where $(C_0, C_1, C_2, C_3, C_4) = (9.5432,\allowbreak -1.6441\times10^{-2},\allowbreak -9.8010\times10^{-3},\allowbreak 4.1664\times10^{-3},\allowbreak  -2.2710 \times 10^{-4})$. We find that eq. (\ref{eq:Treion_fit}) is accurate to within $2\%$ over the range $\vIF = 1\times10^2 - 7\times10^{4}$ km/s.  In \S \ref{sec:implications}, we will describe a procedure for using this fit to model the inhomogeneous thermal history of the IGM. For more detailed applications in which the full dependences on $\spec$ and $\vIF$ are required, we have made publicly available the numerical data for Fig. \ref{fig:Treion_contour}, as well as a simple code for setting up an interpolation.\footnote{\url{cat.ucr.edu}}

\section{Model Expectations}
\label{sec:model_expectations}

Having established the dependence of $\Treion$ on $\vIF$ and $\spec$, we now seek to determine what current models of reionization predict for these quantities.  This will allow us to hone in on the expected values of $\Treion$.  We begin by considering stellar population synthesis modeling of reionization sources in \S \ref{sec:spec}.  We then quantify $\vIF$ from cosmological reionization simulations in \S \ref{sec:velocities}   

\subsection{Spectra of reionization sources}
\label{sec:spec}

In what follows, we assume that Population II stars were the primary sources of ionizing photons during reionization.  We use the Flexible Stellar Population Synthesis (FSPS) code to model the source spectrum \citep{2009ApJ...699..486C, 2010ApJ...712..833C}.  Our fiducial calculations correspond to a single, instantaneous burst of star formation with the initial mass function (IMF) of \citet{2003ApJ...586L.133C}, but we have also explored cases using the \citet{1955ApJ...121..161S} and \citet{2001MNRAS.322..231K} IMFs.\footnote{We adopt lower and upper IMF limits of 0.08$M_{\odot}$ and 120$M_{\odot}$, respectively.}  We find that our results are insensitive to the choice owing to the fact that they differ mainly at sub-solar stellar masses -- a regime which contributes little to the ionizing photon output.  On the other hand, choosing a more top-heavy IMF could significantly impact the spectrum as we discuss below.  We adopt the MESA Isochrones \& Stellar Tracks (MIST; \citealt{2016ApJS..222....8D,2016ApJ...823..102C,2011ApJS..192....3P,2013ApJS..208....4P, 2015ApJS..220...15P}).\footnote{It was necessary to download additional isochrones because FSPS does not (by default) come with isochrone libraries extending to the lowest metallicities considered here ($Z=10^{-3}~Z_{\odot}$). See \url{http://waps.cfa.harvard.edu/MIST}  }  For a detailed comparison of the ionizing spectra of MIST to other models, we refer the reader to \citet{2017ApJ...838..159C}.    

Here we consider the time-integrated spectrum, $\mathcal{S} = \int dt~L_{\nu}(t)$, where $L_{\nu}(t)$ is the specific luminosity at time $t$.  To perform the integral, we sample $L_{\nu}(t)$ at 100 logarithmically spaced times between $t=10^{-4}$ and $500$ Myr.  We have checked that our results are converged with respect to these choices.   Most of the ionizing photons are produced in the first ten million years by massive, short-lived stars.   The time-integrated spectrum should provide a reasonable estimate for the average spectral shape that would be incident on I-fronts if reionization were driven by bursty star formation.  Below, we will discuss additional effects neglected here that would harden the spectrum.    

\begin{figure}
\includegraphics[width=8.7cm]{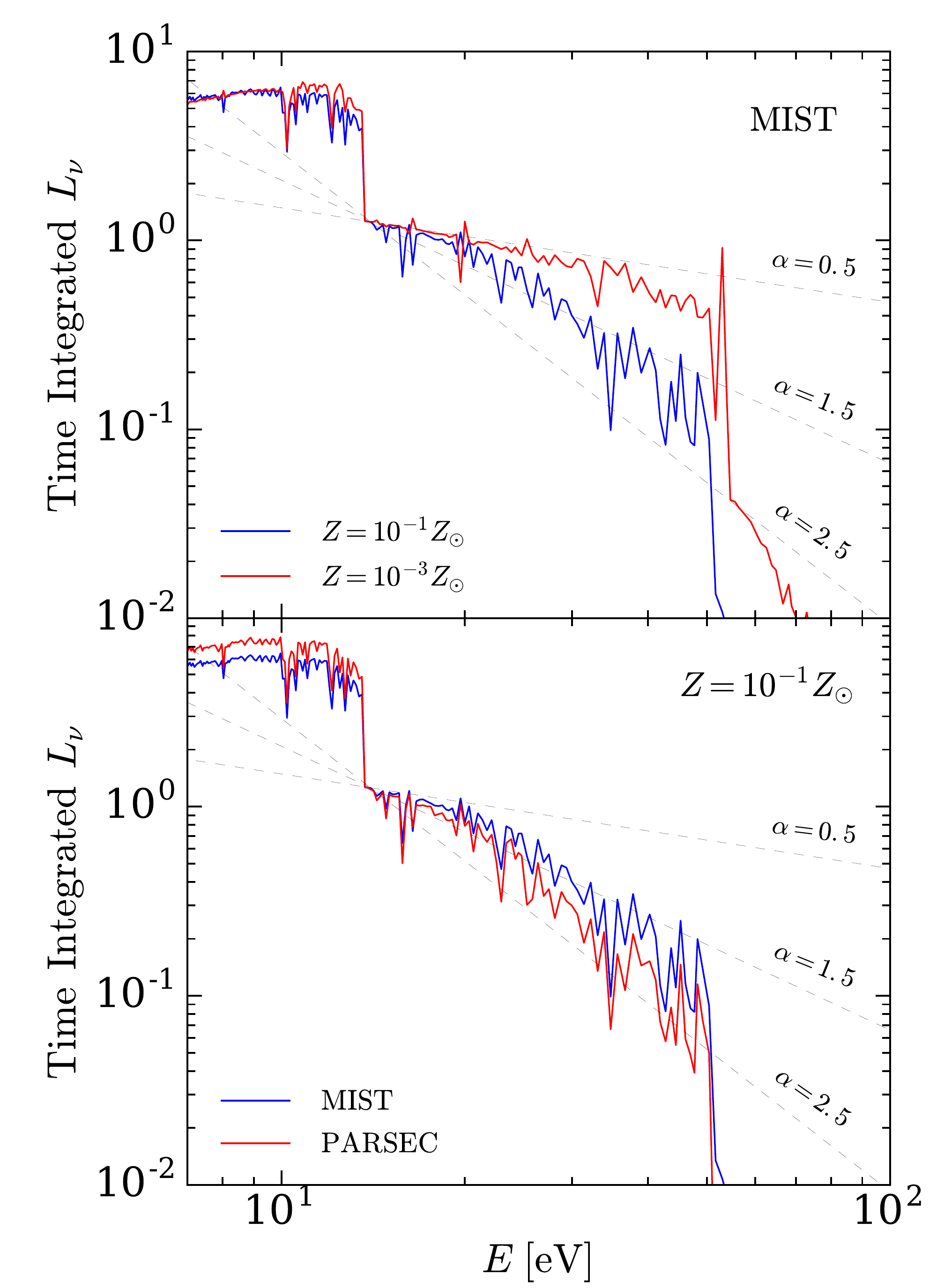}
\caption{Stellar population synthesis models of time-integrated source spectra during reionization.  The units on the $y$-axis are arbitrary. The top panel shows spectra from our fiducial MIST model with $Z=10^{-3}~Z_{\odot}$ and $Z=10^{-1}~Z_{\odot}$.  For reference, the dashed lines correspond to power laws with spectral indices $\alpha = 0.5, 1.5,$ and 2.5.  The bottom panel compares against the PARSEC model at $Z=10^{-1}~Z_{\odot}$.} 
\label{FIG:spectra}
\end{figure}

The red and blue curves in the top panel of Fig. \ref{FIG:spectra} show the integrated spectra for stellar metallicities of $Z=10^{-3}Z_{\odot}$ and $Z=10^{-1} Z_{\odot}$, respectively, roughly bracketing the range of $Z$ found in simulated $z\gtrsim 6$ galaxies with halo masses $M =  10^9 - 10^{12}~\mathrm{M}_\odot$ \citep{2016MNRAS.456.2140M}.  The units on the $y$-axis are arbitrary. (In practice the normalization would be set by the bolometric source luminosity, and our sole focus here is the shape of the spectrum at energies greater than $13.6$ eV.) For reference, the dashed curves correspond to power laws with logarithmic slopes $\alpha = 0.5$, $1.5$, and $2.5$, where we have anchored these curves on the time-integrated spectra near 13.6 eV.\footnote{Here we use $\alpha$ to distinguish this quantity from the spectral index of radiation that is incident on the I-fronts, $\spec$, as the latter may be somewhat lower owing to hardening effects neglected here (see last paragraph of \S \ref{sec:spec}). }  The bottom panel compares against results from a different isochrone and stellar track model at $Z=10^{-1} Z_{\odot}$.  For the red curve, we use the PAdova and TRieste Stellar Evolution Code (PARSEC) model \citep{2012MNRAS.427..127B}, which yields a somewhat softer spectrum.  This difference owes in part to the effects of stellar rotation, which are modeled in MIST \citep[see][]{2017ApJ...838..159C}.     

We can extract {\it effective} spectral indices by matching the mean excess energy per hydrogen ionization between a power-law model and the FSPS spectra. For the MIST models, we estimate effective indices of $\alpha = 0.9(0.7)$ and $1.9(1.4)$ for $Z=10^{-3}Z_{\odot}$ and $Z=10^{-1} Z_{\odot}$, respectively, assuming optically thick (thin) heating. Likewise for PARSEC we obtain $\alpha= 2.3 (1.8)$ for $Z=10^{-1} Z_{\odot}$.  More rigorously, we have also performed  RT runs using the FSPS spectra (see Appendix \ref{sec:convergence}).  For the MIST models, we find that the $\Treion$ values are consistent with $\alpha \approx 0.9$ and 1.75 for the $Z=10^{-3}$ and $10^{-1}~Z_{\odot}$, respectively. For the PARSEC model we find $\alpha \approx 2.25$ for $Z=10^{-1}$.  These values are much closer to the $\alpha$ that we estimated under the assumption of optically thick heating.  Our results indicate that the spectra of metal poor stellar populations are comparable to, or harder than, the spectra of $z=2$ quasars at energies between 1 and 4 Ry. For example, \citet{2015MNRAS.449.4204L} measured $\alpha = 1.70\pm 0.61$ in this regime from their stack of 53 quasars at $z\approx2.4$.  

There are several reasons to suspect that the (galaxy-sourced) ionizing background during reionization may have been somewhat harder than the estimates given here.   First, our calculations neglect the filtering effects of optically thick \HI\ in the ISM of the host galaxy, and within the cosmic web.  Absorption by this gas would have hardened the spectrum of the ionizing radiation as it escaped the galaxy and traveled through the IGM \citep[e.g.][]{1995ApJ...441...18M, 2009ApJ...703.1416F, 2012ApJ...746..125H}. These effects were likely strongest during the last stages of reionization, when the radiation typically had to travel large distances to reach the I-fronts.   Secondly, our calculations neglect the effects of binary star systems.  Mass transfers and mergers between binary companions can extend the period over which ionizing photons are produced by the stellar population, which would harden the time-integrated spectrum \citep{2009MNRAS.400.1019E,2016MNRAS.456..485S}.  Lastly, recent studies have suggested that the IMF in starburst galaxies may be more top-heavy than the IMF assumed here \citep{2005MNRAS.356.1191B,2011MNRAS.415.1647G,2012MNRAS.422.2246M, 2018Natur.558..260Z}.  Most recently, \citet{2018Sci...359...69S} measured a logarithmic slope of $1.90^{+0.37}_{-0.26}$ in the mass range $15-200~\mathrm{M}_{\odot}$, using spectroscopic measurements of the 30 Doradus star forming region in the Large Magellanic Cloud.  (The IMF adopted here has a slope of $2.3$ for $M> M_{\odot}$, and a cutoff of 120 $M_{\odot}$.)  Each of the above effects would work in the direction of making $\spec$ smaller.  Based on the above considerations, we argue that the lower half of Fig. \ref{fig:Treion_contour}, with $\spec \lesssim 1.5$, is likely the most relevant region of parameter space for $\Treion$.  In what follows, we shall adopt $\spec = 1.5$ as our fiducial value, but we note that $\Treion$ is only mildly sensitive to $\spec$ except at the fastest I-front speeds.  In the next section we will find that $\vIF = 10^4$ km/s is close to the upper limit achieved by I-fronts in cosmological simulations, which yields $\Treion = 26,200$ K assuming $\spec=1.5$ (see Fig. \ref{fig:Treion_contour}).  This result varies by $\Delta \Treion = -4,200(+1,600)$ K if we instead assume $\spec=2.5(0.5)$.     

\vspace{5cm}

\subsection{I-front speeds during reionization}
\label{sec:velocities}

In this section, we present calculations of I-front speeds in cosmological simulations of reionization.  

\subsubsection{The SCORCH simulation suite}

We extract I-front speeds from the  {\it Simulations and Constructions of the Reionization of Cosmic Hydrogen} ({SCORCH}) suite \citep{2015ApJ...813...54T, 2016arXiv160503970P, 2017arXiv171204464D}.  In these simulations, the ionizing sources are populated with an abundance matching scheme that connects the UV luminosity of a source to the mass accretion rate of its host halo (see \citealt{2015ApJ...813...54T} for more details).  The reionization simulations were run with the RadHydro code \citep{2004NewA....9..443T, 2007ApJ...671....1T,2008ApJ...689L..81T}.  The Eulerian hydrodynamics module employs non-equilibrium solvers for the ionization and energy equations, and the RT is carried out with adaptive ray tracing based on the HEALPix formalism \citep{2005ApJ...622..759G}.  The radiation spectrum -- discretized into five energy bins above 13.6 eV -- is derived from the stellar population synthesis modeling of \citet{2003MNRAS.344.1000B}.    To reduce computational costs, the simulations adopt a reduced speed of light approximation in which $c_{\mathrm{sim}}$ increases in proportion to the radiation filling factor (i.e. the fraction of cells containing rays), with a minimum value of $0.01c$.  This prescription yields $c_{\mathrm{sim}}/c\approx0.2,0.6$ and 0.9 at volume-weighted ionized fractions of $Q_{\mathrm{HII}} = 0.1, 0.5$ and 0.9, respectively.   We note that these values are larger than the minimum values quoted by previous studies for obtaining reliable I-front speeds.  For example, \citet{2018arXiv180301634D} recently found that $c_{\mathrm{sim}}/c \gtrsim 0.3$ is required to recover I-front speeds reliably throughout reionization, with this condition being relaxed to $c_{\mathrm{sim}}/c \gtrsim 0.05$ during the earliest phases of \HII\ bubble expansion \citep[see also][]{2016ApJ...833...66G}.      

The SCORCH suite consists of three simulations in an $L_{\mathrm{box}} = 50~h^{-1}\Mpc$ box, with $N_{\mathrm{dm}} = N_{\mathrm{gas}}=2048^3$ dark matter particles/gas cells, and $N_{\mathrm{rt}}=512^3$ RT cells.  Three different reionization histories were produced by varying the escape fraction of ionizing radiation, which is parametrized by the redshift-dependent form, $f_{\mathrm{esc}} = A_{\mathrm{esc}}[(1+z)/9]^{\beta_{\mathrm{esc}}}$.  By design, the simulations all yield a Thomson scattering optical depth of $\tau_{\mathrm{es}} \approx 0.06$, and a reionization midpoint of $z\approx 7.5$.  In what follows, we will utilize two of the runs, which we denote using the value adopted for the power-law slope, $\beta_{\mathrm{esc}} = 0$ and $2$, with the latter being our fiducial run.\footnote{Here we use a different notation than in \citet{2017arXiv171204464D}.  Our $\beta_{\mathrm{esc}}$ parameter corresponds to their $a_8$. }  The reionization histories in these runs correspond to the purple and orange curves in Fig. 6 of \citet{2017arXiv171204464D}, respectively.  The end of reionization ($Q_{\mathrm{HII}} = 0.99$) occurs at $z=6.6(5.6)$ in the $\beta_{\mathrm{esc}} = 0(2)$ run, and the duration of reionization -- defined here as the redshift interval between $Q_{\mathrm{HII}}=0.1$ and 0.99 -- is $\Delta z = 3.1(4.8)$.  Both of these models are consistent with the latest CMB and Ly$\alpha$ forest constraints on the timing and duration of reionization \citep{mcgreer15, 2015ApJ...799..177G, 2016A&amp;A...596A.108P}.\footnote{We note that the CMB probes the mass-weighted ionized fraction and not, strictly speaking, the volume-weighted quantities quoted here.  This distinction is not important given the (still large) uncertainties, however.}  

In addition to the standard hydro and RT data outputs, the code also stores the redshift at which each Eulerian cell is reionized. In practice, this is achieved by recording the earliest redshift at which a cell crosses the $50\%$ ionized threshold.  We shall refer to this local quantity as the reionization redshift, $\zreion$, which plays a central roll in our analysis below.    

\subsubsection{Methodology}
\label{sec:vIFmethods}

\begin{figure}
\includegraphics[width=8.8cm]{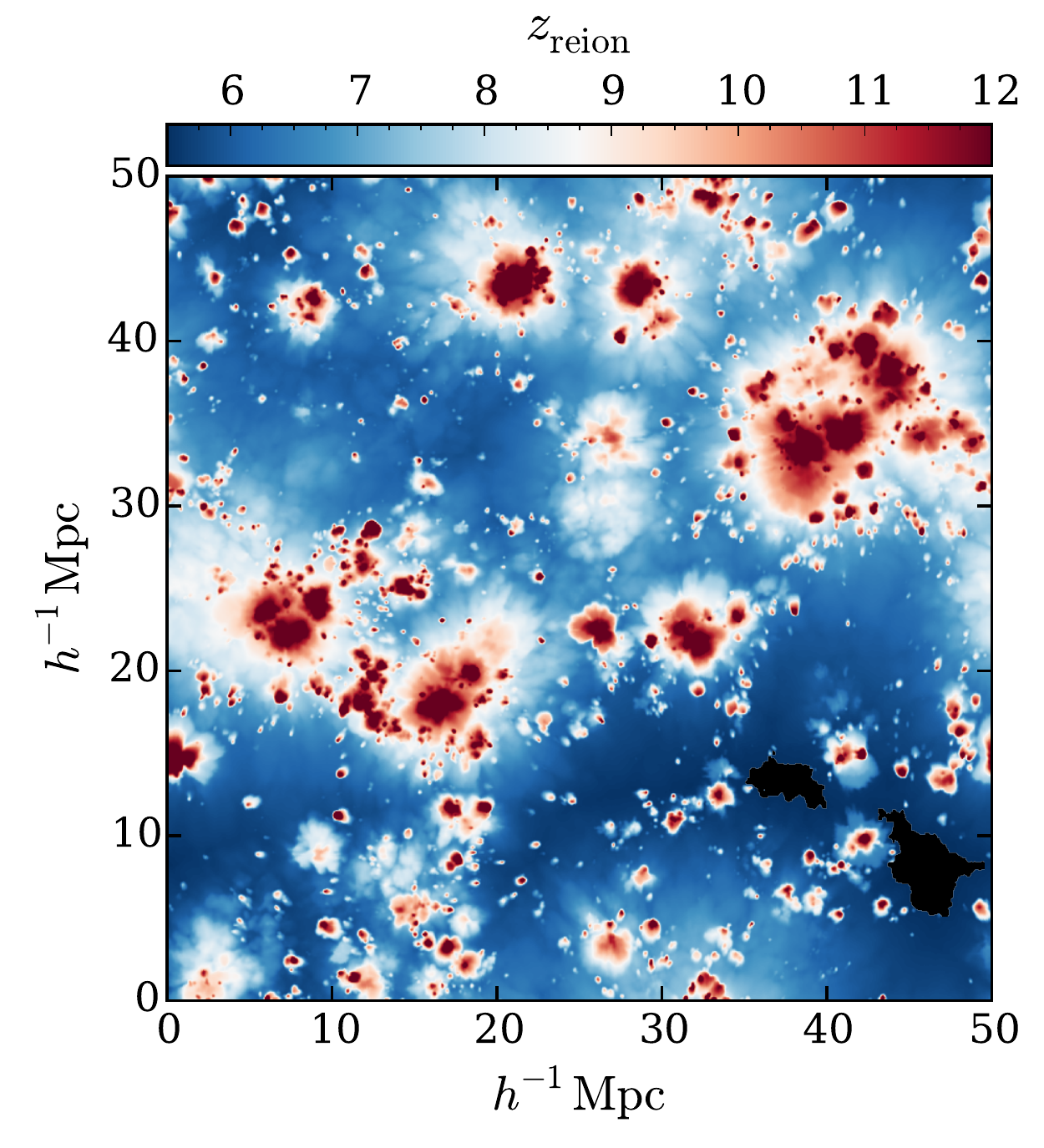}
\includegraphics[width=8.8cm]{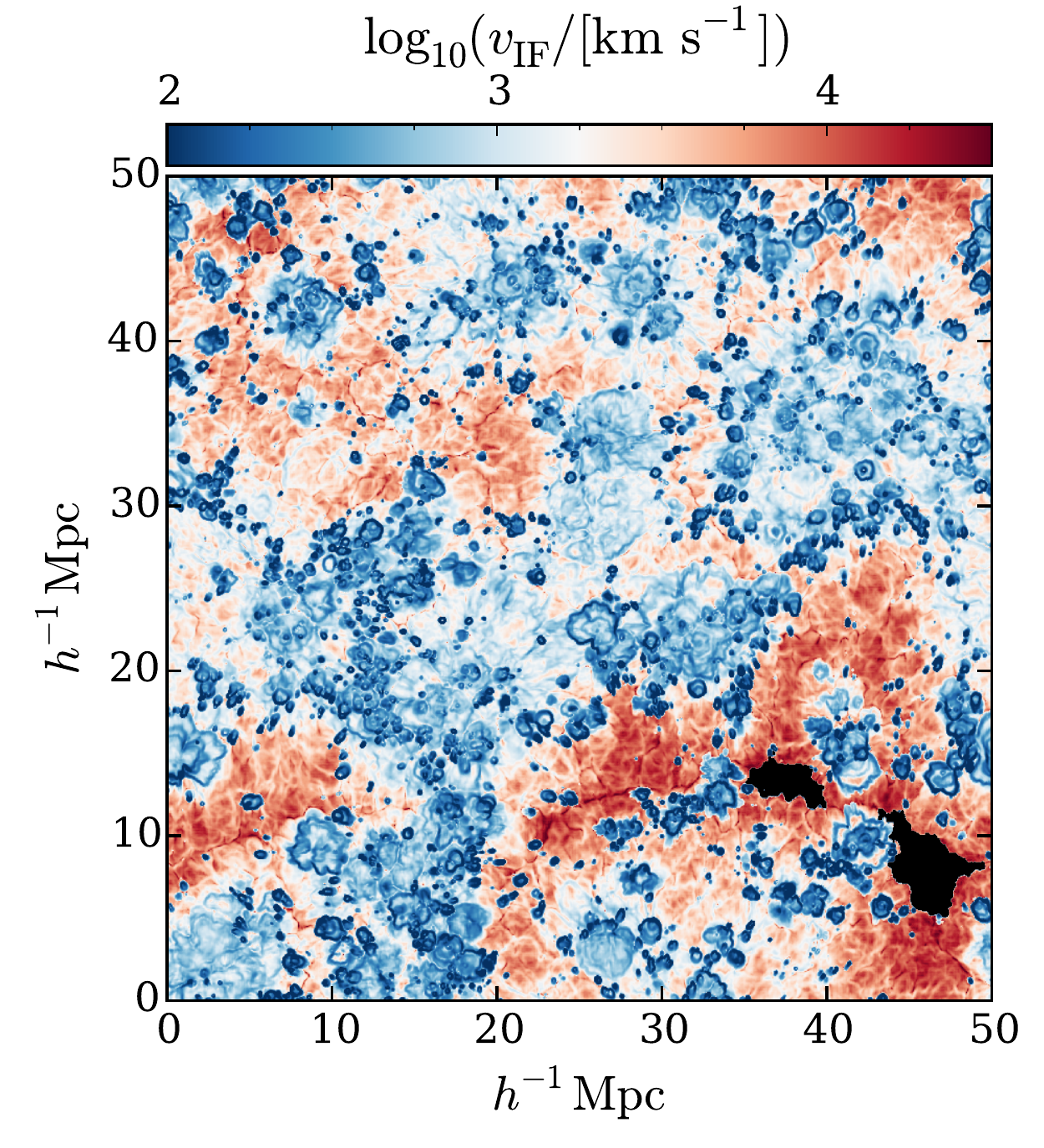}
\caption{Top panel: a slice through the reionization redshift field in our fiducial SCORCH simulation ($\beta_{\mathrm{esc}} = 2$). The $x$- and $y$-axes are in comoving units.  The two black regions in the lower right corner correspond to last remaining patches of neutral gas in the final simulation output at $z=5.5$. Bottom panel: corresponding slice through the field of I-front speeds obtained using the gradient method  (see \S \ref{sec:vIFmethods}).  I-fronts start out moving slowly in the over-dense regions that are reionized first.  I-front speeds increase rapidly as they penetrate into under-dense voids towards the end of reionization.    }
\label{fig:gradientvis}
\end{figure}

We use two independent methods of extracting I-front speeds from the SCORCH simulations.

\smallskip
\noindent
{\bf Gradient Method:} The first method utilizes the reionization redshifts of neighboring cells to compute the local I-front velocity.  Consider a cell at comoving coordinate $\mathbf{x}$ that is reionized at redshift $\zreion = z$.  The proper velocity of the I-front that reionizes the cell can be derived from the local gradient of the $\zreion$ field, 

\begin{equation}
 \mathbf{ \vIF}(\mathbf{x}) = \frac{a}{|\nabla \zreion|} \frac{dz}{dt}~\mathbf{\hat{n}}
 \label{EQ:gradient}
\end{equation}
where the derivatives are with respect to the comoving coordinates $\mathbf{x}$, $a$ is the cosmological scale factor, and $\mathbf{\hat{n}} = \left. (\nabla \zreion/|\nabla \zreion| )\right|_z$ is the unit normal vector to the I-front. (We note that this method of estimating I-front speeds is identical to that of \citealt{2018arXiv180301634D}, which appeared on the archive during the preparation of this manuscript.)  The $\zreion$ fields of the SCORCH simulations are saved at the hydro grid resolution of $N_{\mathrm{gas}}=2048^3$.  As a first step, we smooth the field by convolving with the coordinate-space top-hat function, with smoothing length $L_{\mathrm{box}}/N_{\mathrm{rt}} \approx 98~h^{-1}\kpc$ comoving.\footnote{Alternatively, we also tried re-binning to the RT grid resolution.  We found very similar results between the two methods, with the re-binning method leading to more noise in the $\vIF$ fields.}  This is motivated by the fact that the code can only physically track the propagation of I-fronts at the RT resolution.  We then apply a four-point finite difference to obtain the gradient, $\nabla \zreion$.  

The top and bottom panels of Figure \ref{fig:gradientvis} show a slice through the smoothed $\zreion$ field in our fiducial simulation, and the corresponding $|\mathbf{\vIF}|$ field obtained with equation (\ref{EQ:gradient}), respectively. From these panels a strong correlation between the redshift of reionization and the local I-front speeds is evident.  At the start of reionization, the \HII\ regions expand slowly around the first sources.  The speeds increase as the bubbles grow to encompass more sources, achieving their fastest speeds as they race through the under-dense regions that are reionized last.  We will examine these trends quantitatively below.       

\smallskip
\noindent
{\bf Flux Method:}  As a cross check to the gradient method, we also estimate I-fronts speeds using the flux of ionizing photons at the front boundaries.  We provide details for this method in Appendix \ref{sec:fluxspeeds}.  In summary, the estimator for the I-front speed is

\begin{equation}
\vIF = \frac{c F }{F+c  n_{\mathrm{H}}(1+\chi)},
\label{EQ:fluxvel}
\end{equation}
{where $F$ is the number flux of ionizing photons at the front boundary}, and the factor $1+\chi = 1 + n_{\mathrm{He}}/n_\mathrm{H} \approx 1.08$ accounts for singly ionized Helium.  We use the $\zreion$ fields from the simulations to identify optically thin RT cells near the I-front boundaries at a given snapshot in time.  For a given boundary cell, we estimate $F$ by counting up the number of ionizing photons within the cell.  The local hydrogen number density, $n_{\mathrm{H}}$, is obtained by smoothing the hydro density field to the RT resolution.  The flux method overestimates $\vIF$ because it assumes that all of the cell's photons propagate in a direction that is normal to the I-front.  However, the worst that the local $\vIF$ can be overestimated under this assumption is a factor of $2$ if the radiation is impinging uniformly from all directions onto the plane of the I-front.  In realistic situations the radiation is likely more directional such that the flux method provides a closer estimate to the true $\vIF$.  

Lastly, we note the possibility of using equation (\ref{EQ:fluxvel}), in combination with the fit of equation (\ref{eq:Treion_fit}), to lay down $\Treion$ values ``on-the-fly" in cosmological RT simulations that would otherwise be unconverged in $\Treion$ (e.g. monochromatic simulations).         

\subsubsection{Results}

\begin{figure}
\includegraphics[width=8.5cm]{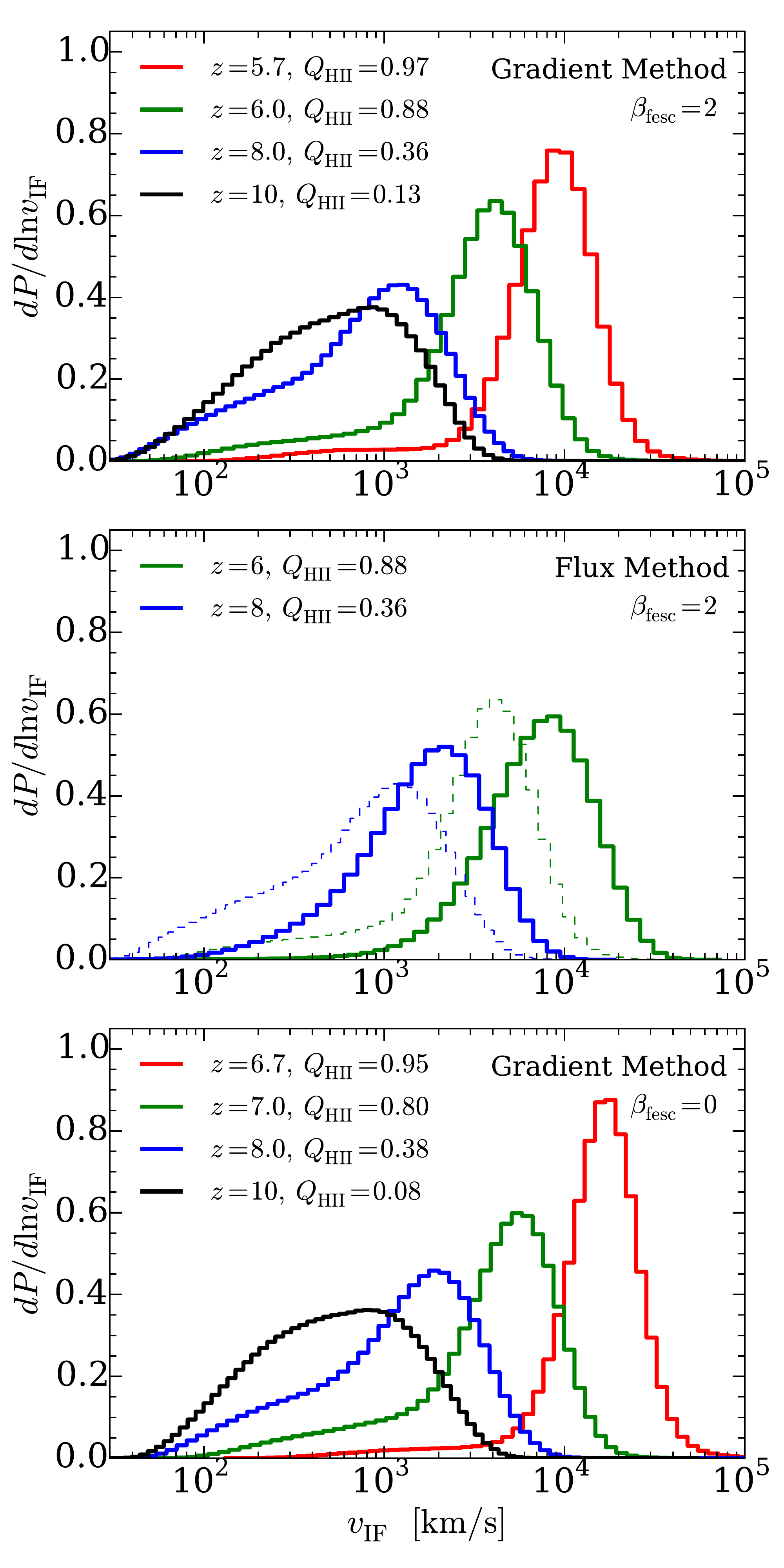}
\caption{Distribution of I-front speeds in the SCORCH reionization simulations.  From left to right, the distributions correspond to progressively later snapshots in time. The top and middle panels compare speeds in our fiducial simulation ($\beta_{\mathrm{esc}} = 2$) obtained from the gradient and flux methods, respectively. (For ease of comparison, the gradient results are represented by thin dashed curves in the middle panel.)  The bottom panel shows speeds in the $\beta_{\mathrm{esc}} = 0$ simulation, in which the duration of reionization is shorter.  I-fronts move faster in models with a shorter duration of reionization, leading to hotter post-I-front temperatures.     }
\label{fig:vIFdists}
\end{figure}

We begin by comparing the gradient and flux methods.  The top panel of Fig. \ref{fig:vIFdists} shows the probability distributions of $\vIF$ at four snapshots in time using the gradient method on our fiducial simulation.  The solid histograms in the middle panel show the flux method results at two of the redshifts for which full simulation outputs were saved.  For ease of comparison, the thin/dashed histograms show the corresponding gradient method distributions, reproduced from the top panel.  We note that the results disagree by a factor of $\approx$ 2, with the flux method yielding faster speeds.  While this factor apparently corresponds to the maximum possible amount that the flux method can overestimate $\vIF$ (as described above), we argue that this discrepancy is unlikely to result solely from the flux method limitations.  It may indicate that the gradient method is also underestimating I-front speeds, perhaps due to noise in the $\zreion$ fields.  However, the flux method is of limited utility because it can only be applied to redshifts at which we have full RT outputs.  Since the gradient method allows us to obtain $\vIF$ at all times using just the $\zreion$ field, we will adopt its slower speeds as our fiducial results, which results in somewhat lower $\Treion$. 

The top and bottom panels of Fig. \ref{fig:vIFdists} compare the speeds in the $\beta_{\mathrm{esc}} = 2$ and $0$ simulations, where the duration of reionization is shorter in the latter.  Intuitively, the shorter the duration of reionization, the faster the I-fronts must move.  In both cases, the distributions are broad at early times ($Q_{\mathrm{HII}}\sim 10\%$), spanning two orders of magnitude from $\sim 50$ to $2 \times 10^3~\mathrm{km/s}$.   During this phase of reionization, \HII\ bubbles are expanding from individual (or few) sources that are typically embedded in over-dense regions of the universe.  The slowest speeds in the distribution correspond to I-fronts that are retarded either by dense regions surrounding the sources, or by their episodic star formation histories.   As reionization progresses, the \HII\ regions begin to encompass more sources, such that the flux of ionizing photons at the front boundaries increases rapidly.   The I-fronts break free from the over-dense regions and expand quickly through the voids.  Thus we observe a strong evolution in the I-front speeds, with $\vIF$ reaching $\sim 10^4~\mathrm{km/s}$ prior to overlap.  These results are broadly consistent with the findings of \citet{2018arXiv180301634D}, who also found a progression towards rapid speeds near overlap.  In the next section, we will translate these I-front speeds to post-I-front temperatures.  

\begin{figure}
\includegraphics[width=8.5cm]{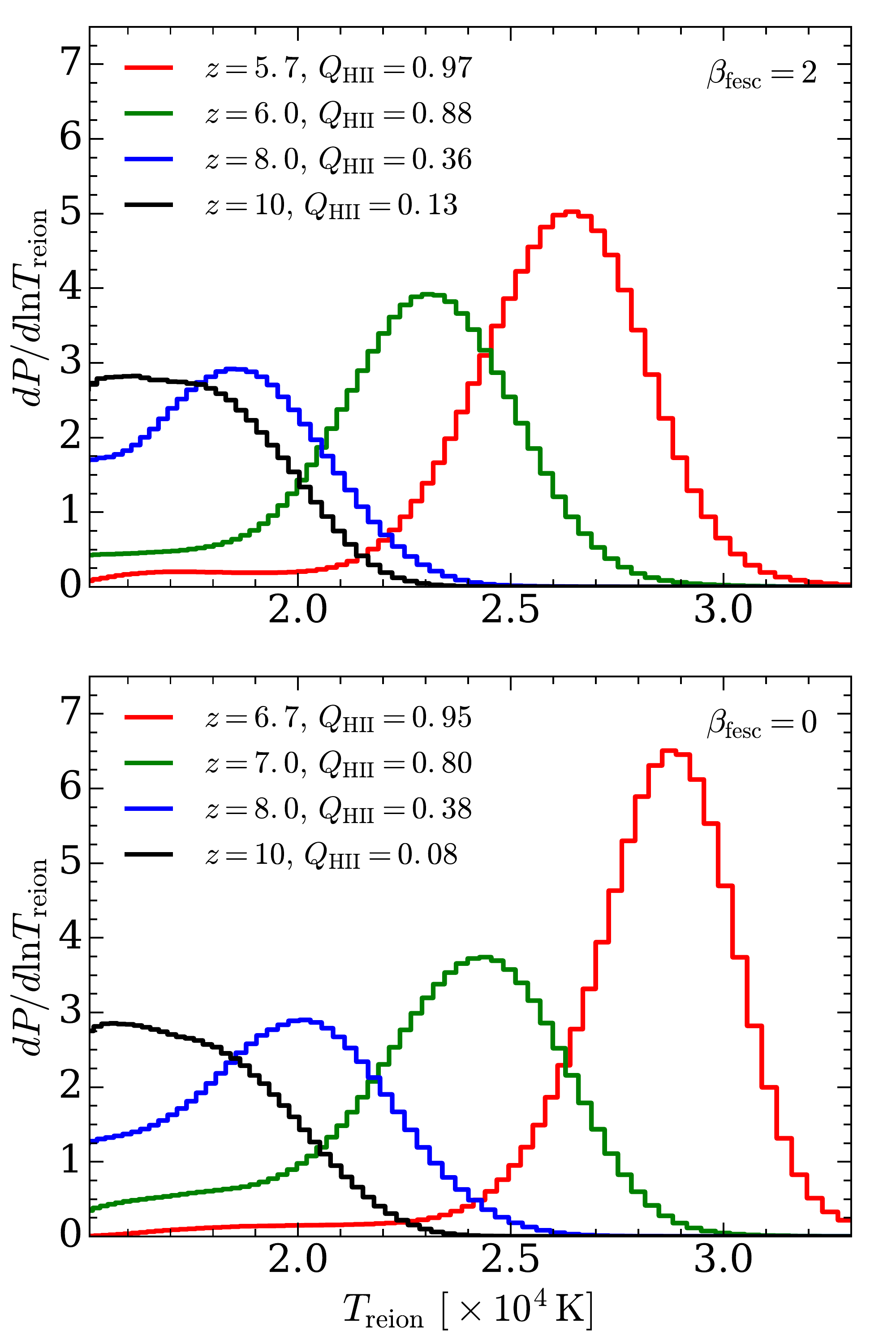}
\caption{Models for the distribution of post-I-front temperatures during reionization.  We use the the fit of equation (\ref{eq:Treion_fit}) to map the velocity distributions in Fig. \ref{fig:vIFdists} to $\Treion$ values.  Velocities were obtained using the gradient method and we assume a spectral index of $\spec=1.5$.  The post-I-front temperatures become hotter throughout reionization, with $\Treion \approx 25,000 - 30,000$ K near the end of this process.}
\label{fig:Treiondists}
\end{figure}

\begin{figure*}
\centerline{
\includegraphics[width=8.0cm]{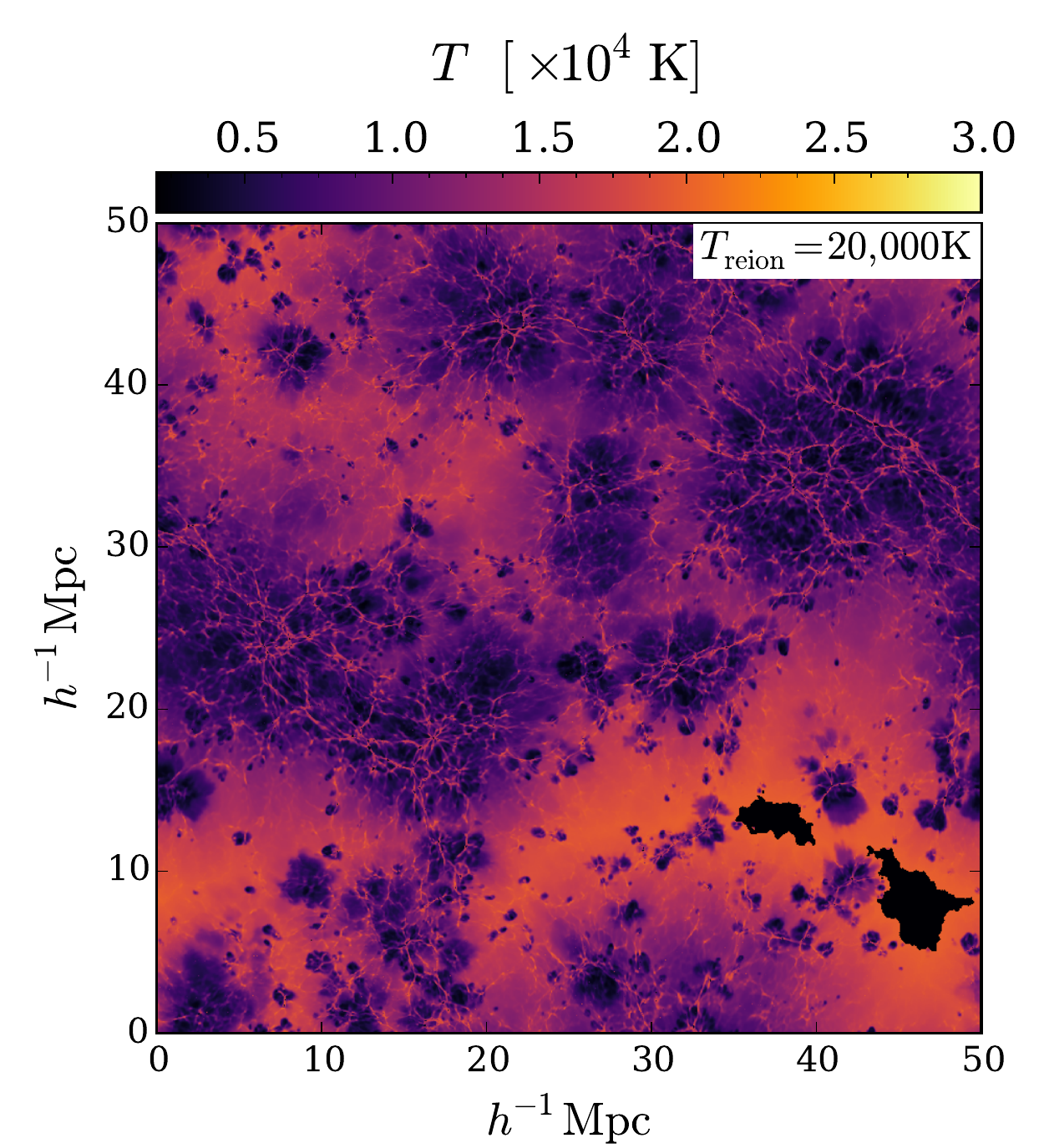}
\includegraphics[width=8.0cm]{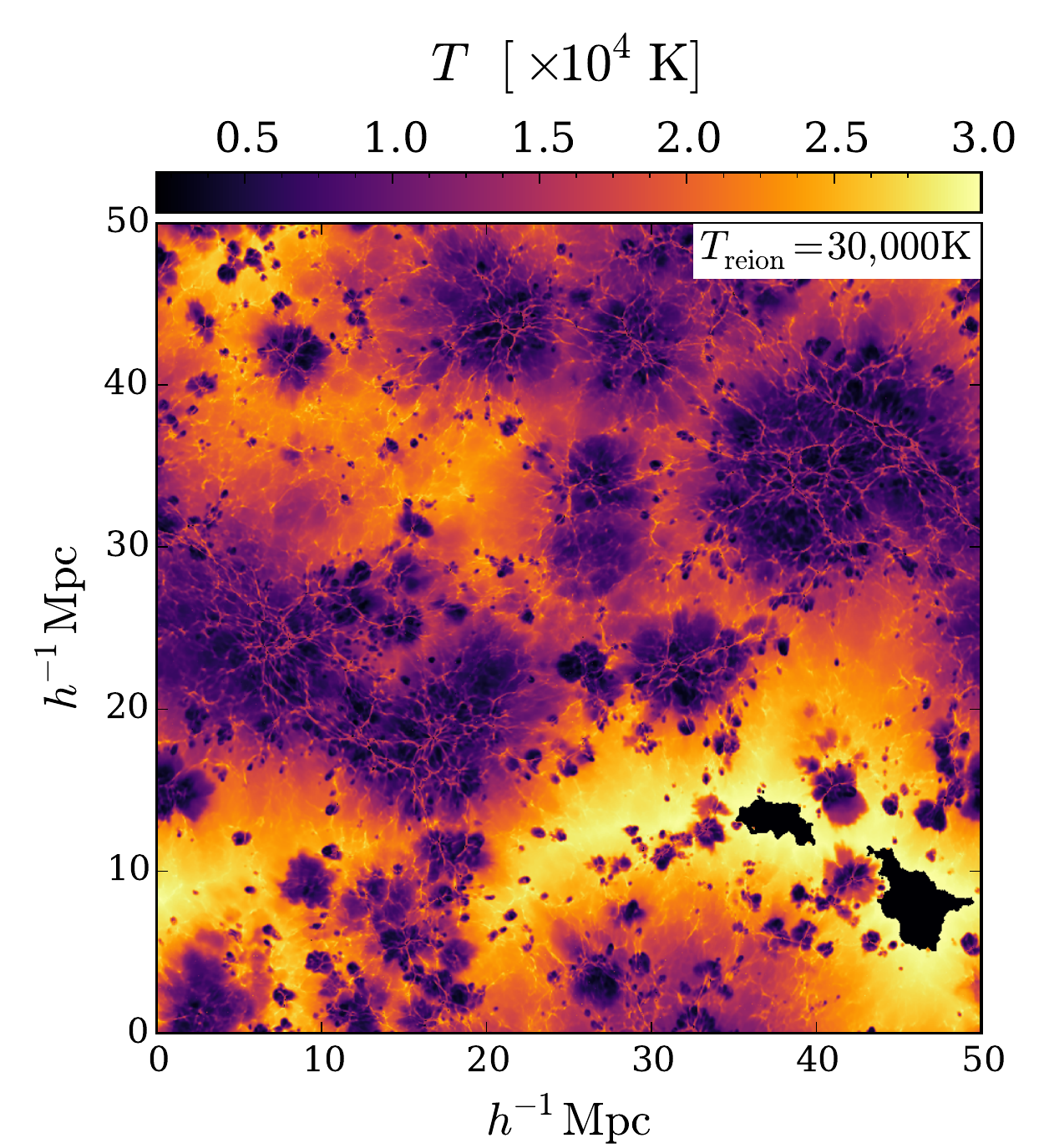}}
\centerline{
\includegraphics[width=8.0cm]{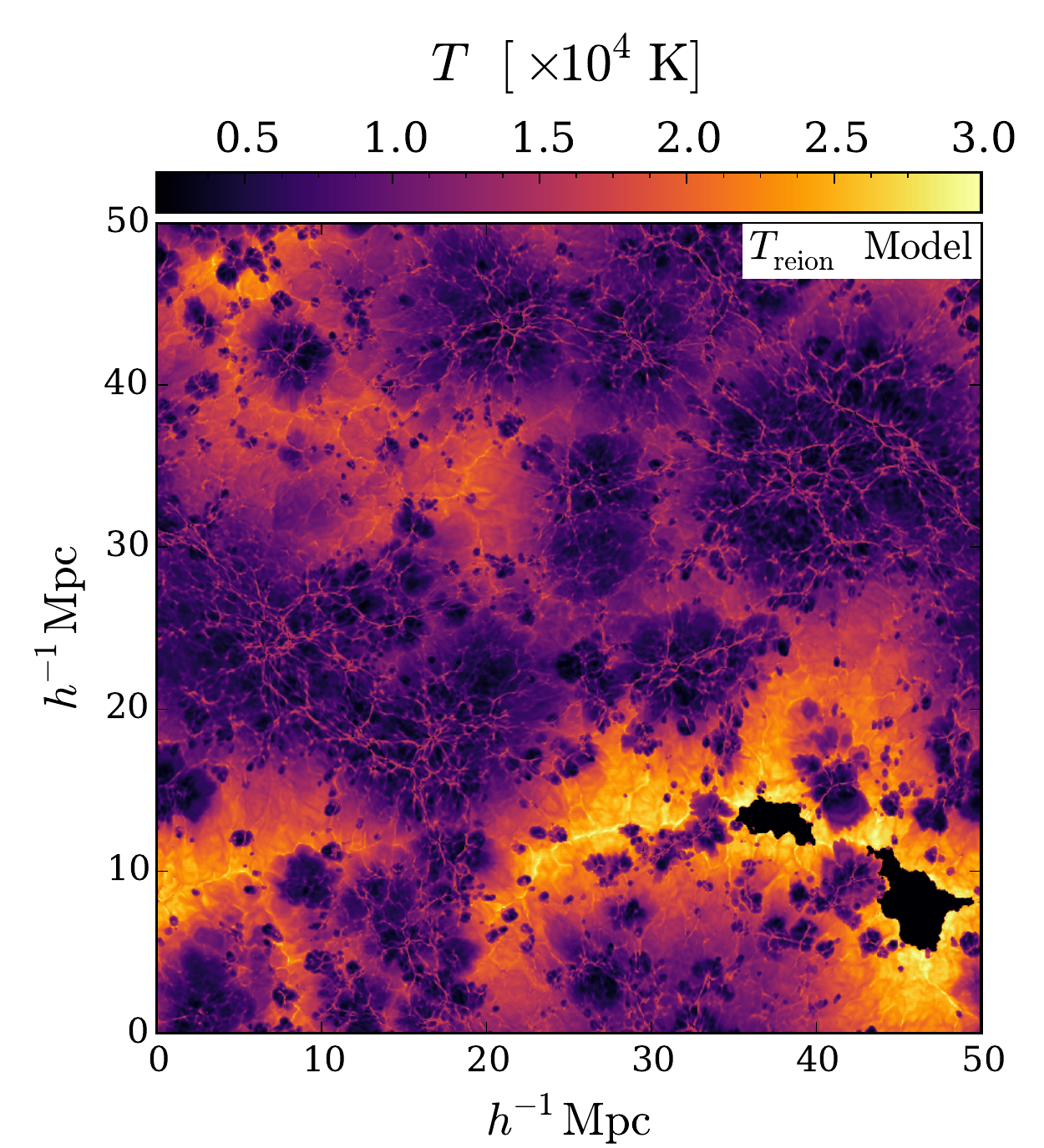}
\includegraphics[width=8.0cm]{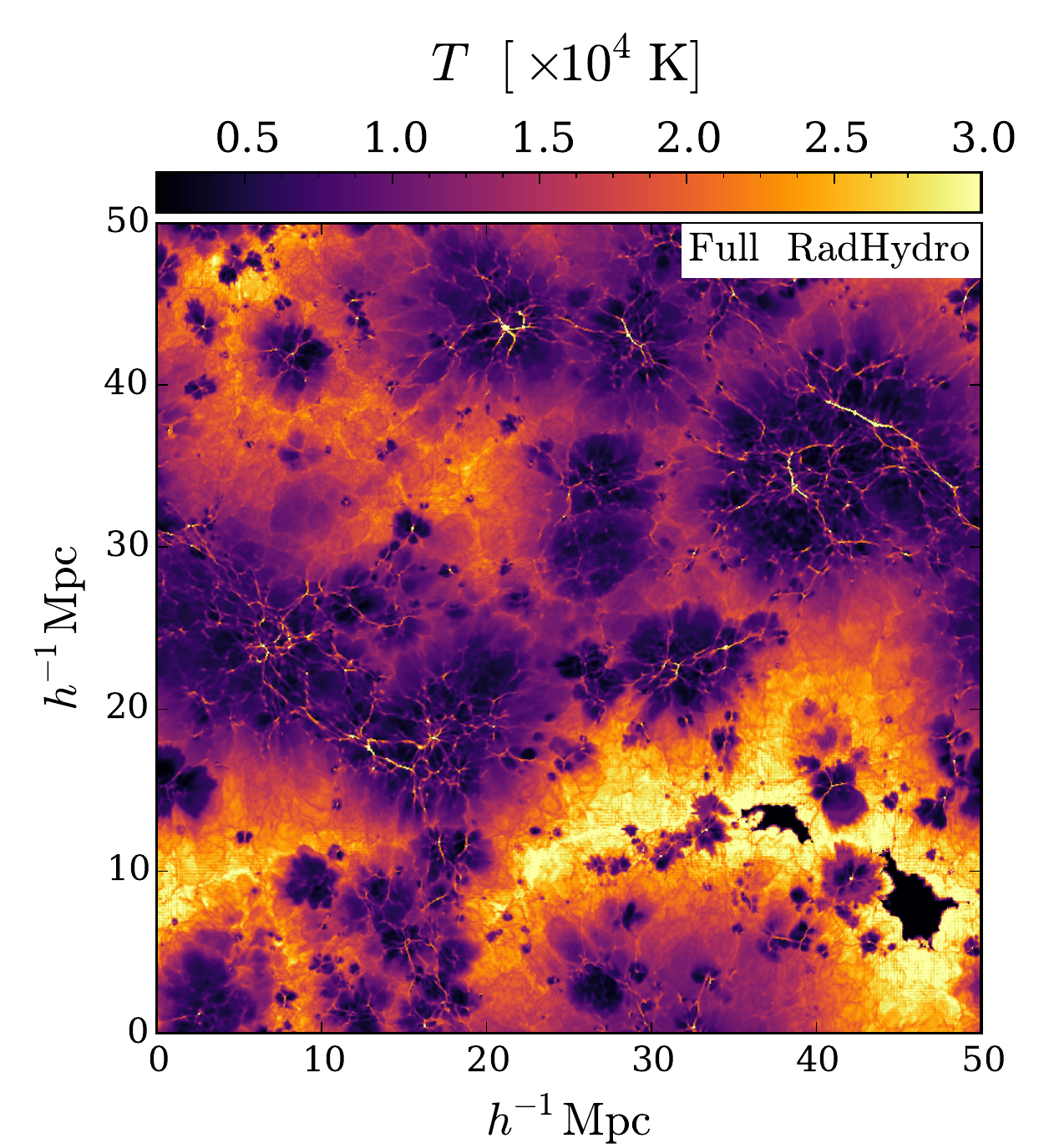}
}
\caption{Slices through the temperature field at $z=5.5$ in three models for $\Treion$ (top-left, top-right, and bottom-left) compared to the temperatures in the full Radhydro simulation (bottom-rght).  Here we use results from our fiducial SCORCH simulation with $\beta_{\mathrm{esc}} =2$.  The panels labeled $\Treion=20,000$ and 30,000 K assume a fixed $\Treion$, while the panel labeled ``$\Treion$ Model" corresponds to our new model for laying down $\Treion$ based on I-front velocities.  The two black patches in the lower right-hand corners of the panels correspond to last remaining patches of neutral gas.   }
\label{fig:Tslices}
\end{figure*}

\section{Thermal history of the IGM}
\label{sec:implications}

In this section, we synthesize our results into a model for exploring the impact of $\Treion$ on the thermal history of the IGM.  Let us begin by describing a simple prescription for laying down $\Treion$ values for any given $\zreion$ field. The first step is to apply the gradient method of \S \ref{sec:velocities} to compute $|\mathbf{\vIF}|$ at each location in the $\zreion$ field.  Then, the fit of equation (\ref{eq:Treion_fit}) can be used to translate these speeds to post-I-front temperatures at each location.  In Figure \ref{fig:Treiondists} we apply this procedure to the SCORCH $\zreion$ fields to obtain $\Treion$ distributions.  Each histogram corresponds to a particular time during reionization, with redshift decreasing towards the right.  The distributions are broad throughout, but the temperatures evolve significantly as reionization progresses.  Post-I-front temperatures are typically around $17,000$ K during the early phases of reionization ($Q_{\mathrm{HII}} = 0.1$), but the mean values reach $\Treion = 26,000~(29,000)$ K near the end of this process in the $\beta_{\mathrm{sec}} = 2~(0)$ models.  Models with a shorter duration of reionization lead to hotter temperatures, as the I-fronts must traverse the same volume in a shorter time frame.      

Next, we evolve the temperatures in time to explore what our results imply for the thermal history of the IGM.  We adopt an approach similar to that of \citet{2017arXiv170808927D} and \citet{2016MNRAS.460.1885U}.   In this simplified model, the temperature of each gas parcel evolves according to 

\begin{equation}
\frac{dT}{dt} = -2 H T + \frac{2 T}{3 \Delta}\frac{d \Delta}{d t} - \frac{T}{n_{\mathrm{tot}}} \frac{d n_{\mathrm{tot}}}{dt} + \frac{2}{3 k_\mathrm{B} n_{\mathrm{tot}}} \frac{d Q}{dt},
\label{eq:Tevol}
\end{equation}
where $H$ is the Hubble parameter.  The $dQ/dt$ term includes \HI\ and \HeI\ photoheating and all of the relevant cooling processes for ionized gas of primordial composition \citep{1997MNRAS.292...27H}.  \citet{2016MNRAS.456...47M} showed that the temperature of a gas parcel at a given density will be nearly the same for any reasonable model of its prior density evolution.  For simplicity, we adopt the Zel'dovich pancake approximation for the adiabatic compression/expansion term in which $\Delta(a)=\left[ 1- \lambda G(a) \right]^{-1}$, where $G(a)$ is the linear growth factor, and the constant $\lambda$ is adjusted to match the simulation densities at the redshift of interest. (Here we shall consider $z=5.5$.) For each cell in the SCORCH $\zreion$ fields, we set $T=\Treion$ at the appropriate redshift and solve the differential equation numerically to get $T$ at a later time.  There are two important caveats to this approach. First, equation (\ref{eq:Tevol}) does not account for shock heating by collapsing structures, which will become evident when we compare our results against the full SCORCH simulation results below. Second, it is incorrect -- strictly speaking -- to apply this Lagrangian equation to our Eulerian gas cells.  However, we will see that this simple and computationally inexpensive approach reproduces the large-scale structure of the temperature field. 

Using our fiducial SCORCH run, Fig. \ref{fig:Tslices} shows 2D slices of the temperature fields for three models at $z=5.5$.  We note that reionization ends at $z\approx 5.5$ in this simulation, so these slices represent snapshots of the temperature field at the end of reionization.  Indeed, the black islands correspond to the last remaining patches of neutral gas. The top-left and top-right panels show models with spatially uniform $\Treion = 20,000$ and $30,000$ K, respectively.  The bottom-left corresponds to our new model for laying down $\Treion$.  For comparison, the bottom right panel shows the temperature field extracted directly from the full RadHydro simulation data.  First we note the higher temperatures in the filaments due to shock heating, which is not included in our simple models.  It is also evident that the model with spatially uniform $\Treion=20,000$ K does not capture the hottest post-I-front temperatures that are imprinted near the end of reionization.  Even compared to our full $\Treion$ model (bottom-left), the temperatures of recently reionized patches are somewhat hotter in the full RadHydro simulation (bottom-right).  This may be a symptom of the gradient method underestimating $\vIF$, or spectral filtering, which is not included in our model. 

\begin{figure}
\includegraphics[width=8.7cm]{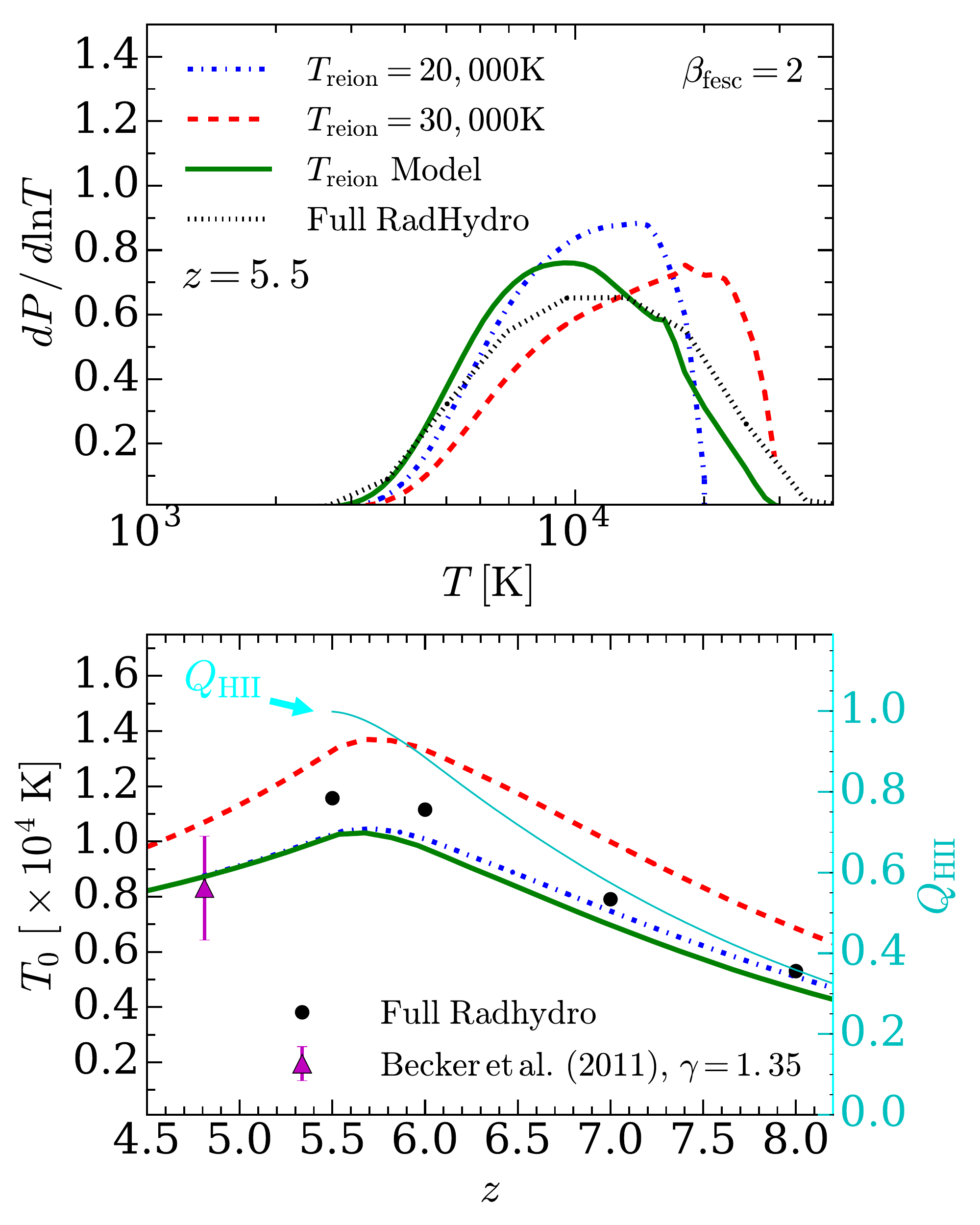}
\caption{Effects of reionization on the thermal history of the IGM. Top panel: Distribution of temperatures (across all densities) at $z=5.5$ in three models compared to the full Radhydro simulation results.  The models correspond to those in Fig. \ref{fig:Tslices}. Bottom panel: Temperature at the mean gas density ($T_0$).   The triangular data point corresponds to the measurement of \citet{2011MNRAS.410.1096B}.  For reference, the thin/cyan curve shows the volume filling fraction of ionized hydrogen, $Q_{\mathrm{HII}}$ (right axis).}
\label{fig:dP_dT}
\end{figure}

\begin{figure}
\includegraphics[width=8.7cm]{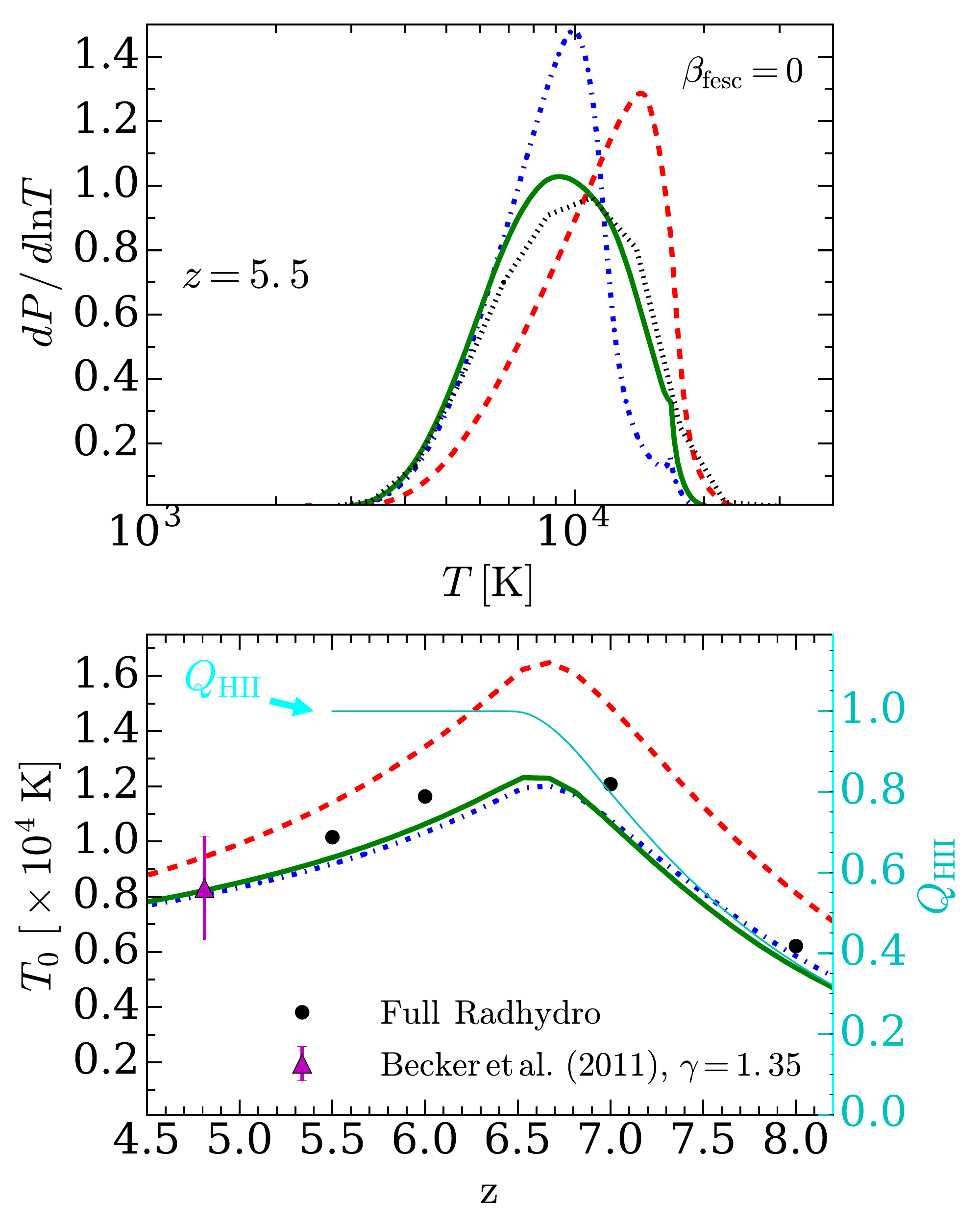}
\caption{Same as in Fig. \ref{fig:dP_dT}, but for the $\beta_{\mathrm{esc}}=0$ simulation in which the reionization epoch is shorter and ends earlier. }
\label{fig:dP_dT_B0}
\end{figure}

For a more quantitative look at these temperatures, the top panel of Fig. \ref{fig:dP_dT} shows the temperature distributions at $z=5.5$ (across all densities). As noted above, the uniform $\Treion=20,000$ K model misses the high-temperature tail of the distribution, and the full RadHydro distribution is somewhat wider than that of our model. The bottom panel examines the evolution of the temperature at the mean density of the universe in the three models.  To make contact with observational results, we have included the $z\approx 4.8$ temperature measurement of \citet{2011MNRAS.410.1096B}, extrapolated to the mean density using a temperature-density relation of the form $T=T_0 \Delta^{\gamma-1}$, with $\gamma =1.35$. (We note that, at this redshift, the measured $T_0$ is not sensitive to the choice of $\gamma$).  As expected, the uniform $\Treion =30,000$ K model yields a significantly hotter mean temperature compared to the other models, but at $z\approx 4.8$ even this model is statistically consistent with the measurement.   The top panel shows that our model for $\Treion$ yields a wider distribution of post-reionization temperatures compared to the uniform $\Treion=20,000$ K case.  However, the bottom panel shows that the $T_0$ evolution for these two models is similar.  This is because much of the volume in our model is, in fact, reionized to temperatures $\Treion \approx 20,000$ K, with the hotter values of $\Treion=25,000 - 30,000$ K relegated to the fraction of the volume that is reionized late (providing the hot tail of the distribution).  

Figure \ref{fig:dP_dT_B0} shows the corresponding results from the $\beta_{\mathrm{fesc}} = 0$ model, which completes reionization earlier at $z\approx 6.6$. Although hotter temperatures are reached at the end of reionization in this model, the gas is somewhat colder by $z=5.5$.  The distribution of temperatures is also narrower, illustrating that the amplitude of temperature fluctuations is sensitive to the timing of overlap.  

The results of this section indicate that current models of reionization yield $\Treion$ as high as $25,000 - 30,000$ K near the end of this process -- a consequence of the fast I-front speeds that are achieved in these models.  The range of maximum temperatures is narrower and lower, $\Treion = 23,000 - 25,000$ K, if the softest model considered in \S \ref{sec:spec} (with $\spec \approx 2.25$) is more representative of the sources that drove the end of reionization.  We note that our $\Treion$ values are generally higher than those obtained in one-zone approximations of the reionization heating (see e.g. Appendix A of \citealt{2018arXiv180104931P}).  Such calculations do not capture the heating/cooling structure within I-fronts, which is crucial for obtaining accurate temperatures.  Our maximum $\Treion$ values are also higher than those found in the cosmological RT simulations of \citet{2018arXiv180500099F}.  Their lower temperatures -- which never exceed 20,000 K -- may owe to a combination of low RT spatial resolution (relative to the I-front widths) and frequency binning (K. Finlator, personal communication).  It is also possible that the smaller box size used in \citet{2018arXiv180500099F} ($12 h^{-1}~\Mpc$) produces lower fluxes at the I-fronts, and therefore lower front velocities, particularly towards the end of reionization.            

If reionization ended around $z=6$, the hottest patches in our models could persist long enough to be detected in the $z> 5$ Ly$\alpha$ forest, providing a potential target for future observational studies.  An important caveat here is that it is uncertain whether I-front speeds ever achieved the required speeds of $\sim10^4$ km/s in actuality. The maximum speeds depend on the poorly understood sinks of ionizing photons at these epochs.  Near the end of reionization, absorptions by self-shielding gas in the cosmic web regulate the expansion rate of \HII\ regions \citep{2000ApJ...530....1M, 2005MNRAS.363.1031F}.  Current RT simulations probably have yet to achieve the spatial resolutions needed to capture this regulation effect fully, so the maximum speeds may have been slower than we find here.  We emphasize that Fig. \ref{fig:Treion_contour} provides the connection between I-front speeds and post-I-front temperatures irrespective of the reionization model.  For example, if the maximum I-front speeds are revised downwards from $10^{4}$ to $5\times 10^3$ km/s in future simulations, Fig. \ref{fig:Treion_contour} shows that the maximum $\Treion$ would come down by $\approx 3,000$ K.  

Lastly, we comment on the interpretation of the $160$ comoving Mpc dark Ly$\alpha$ trough towards quasar ULAS J0148$+$0600 \citep{2015MNRAS.447.3402B}.  Previous studies have invoked large ionizing background \citep{2016MNRAS.460.1328D, 2017MNRAS.465.3429C} or temperature fluctuations \citep{2015ApJ...813L..38D} to explain the existence of this trough.  Recently, \cite{2018arXiv180308932B} conducted a narrow-band survey towards this sightline and found a significant under-density of Ly$\alpha$ emitters extending radially out to $30$ comoving $\Mpc$ from the trough.  These results indicate that the sightline intersects a cosmic void and that its Ly$\alpha$ opacity likely owes to a highly suppressed local ionizing background \citep{2017arXiv170808927D}.   In this case, the associated void cannot be too hot without yielding detectable Ly$\alpha$ transmission in conflict with observed upper limits \citep{2015ApJ...813L..38D}.  There are two ways to reconcile these observations with the models presented here: either the trough was reionized much earlier than $z=6$, or it was reionized rather slowly compared to the voids in our simulations.  We note that the large-scale fluctuations that would be required to explain these observations are not captured in any reionization simulation to date, including the ones that we used here to study I-front speeds. (Nor should we expect them to be, since the fluctuations occur on scales similar to or larger than the typical box sizes of the simulations.)  The currently favored model for the large-scale fluctuations requires that the mean free path be a factor of $\gtrsim 2$ shorter than extrapolations of measurements at $z\lesssim5.2$ \citep{2016MNRAS.460.1328D, 2018MNRAS.473..560D}.  If confirmed, this would imply that absorptions played an important role in regulating the reionization process.  Future studies should investigate the parameter space for reionization that is consistent with this scenario.

\section{Conclusion}
\label{sec:conclusion}

We have presented a study of post-I-front temperatures during reionization.  We used a suite of high-resolution RT simulations to quantify the dependence of $\Treion$ on the I-front speed and the spectrum of incident radiation.  We found that post-I-front temperatures are only mildly sensitive to the spectral index of the incident radiation over most of the parameter space, with $\Treion$ set primarily by the local I-front speeds.  The results of our parameter space study can be used to map I-front speeds to $\Treion$. 
 
We then measured I-front speeds in cosmological RT simulations to determine what current models of reionization predict for $\Treion$.  The distribution of speeds is broad during the early phases of reionization, with values ranging from 50 to $2 \times 10^3~\mathrm{km/s}$.  However, $\vIF$ increases and the distribution narrows with time such that $\vIF \sim 10^4~\mathrm{km/s}$ near overlap.  Mapping these velocities to temperatures yields $\Treion = 17,000-22,000$ K during the first half of reionization, but hotter temperatures of $\Treion = 25,000 - 30,000$ K are reached near overlap.  A shorter duration of reionization generally implies hotter temperatures, since the I-fronts must move at faster speeds.  

If reionization ended near $z=6$, our models suggest that hot, recently reionized gas may be observable in high-resolution quasar absorption spectra.  In addition to being hot, these regions should exhibit a mildly inverted temperature-density relation, reflecting the slower(faster) speeds at which I-fronts move through over(under)-densities.  Such regions provide a potential target for future studies pursuing signatures of reionization in high-$z$ quasar absorption spectra.   It is worth noting that a lack of evidence for hot regions would also lead to important insights on reionization.  For example, their absence may indicate that reionization ended significantly earlier than $z=6$.  It is also possible that I-fronts moved at slower speeds near overlap than is predicted by contemporary models.  Since the cold, pre-reionization gas clumps on scales below the resolution limits of current simulations, the poorly understood sinks of ionizing photons may not be well captured in our models.  In this case, the lack of hot gas would ultimately provide observational insight into the role of sinks in setting the speed limit for I-fronts.      

Modeling the signatures of reionization in quasar absorption spectra requires a confluence of ionizing background and temperature effects, where accurate post-I-front temperatures are a key ingredient for the latter. The results of our parameter space study can be applied to future models of the thermal history. 

\acknowledgments

A.D. thanks Vid Ir\v{s}i\v{c}, Nell Byler, George Becker, Kristian Finlator, and Brian Siana, for useful discussions and comments on a draft of this manuscript.  A.D. acknowledges HST grant HST-AR-15013.005-A and NSF XSEDE allocation TG-AST150004.  M.M. acknowledges NSF grants AST 1514734 and AST 1614439, NASA ATP award NNX17AH68G, the Alfred P. Sloan foundation, and NSF XSEDE allocation TG-AST140087. H.T. acknowledges HST grant HST-AR-15013.005-A. 

\bibliographystyle{apj}
\bibliography{./Bibliography}

\begin{thebibliography}{}
\expandafter\ifx\csname natexlab\endcsname\relax\def\natexlab#1{#1}\fi

\bibitem[{{Ba{\~n}ados} {et~al.}(2018){Ba{\~n}ados}, {Venemans},
  {Mazzucchelli}, {Farina}, {Walter}, {Wang}, {Decarli}, {Stern}, {Fan},
  {Davies}, {Hennawi}, {Simcoe}, {Turner}, {Rix}, {Yang}, {Kelson}, {Rudie}, \&
  {Winters}}]{2018Natur.553..473B}
{Ba{\~n}ados}, E., {Venemans}, B.~P., {Mazzucchelli}, C., {et~al.} 2018, \nat,
  553, 473

\bibitem[{{Banks}(1966)}]{1966P&amp;SS...14.1105B}
{Banks}, P. 1966, \planss, 14, 1105

\bibitem[{{Baugh} {et~al.}(2005){Baugh}, {Lacey}, {Frenk}, {Granato}, {Silva},
  {Bressan}, {Benson}, \& {Cole}}]{2005MNRAS.356.1191B}
{Baugh}, C.~M., {Lacey}, C.~G., {Frenk}, C.~S., {et~al.} 2005, \mnras, 356,
  1191

\bibitem[{{Becker} \& {Bolton}(2013)}]{2013MNRAS.436.1023B}
{Becker}, G.~D., \& {Bolton}, J.~S. 2013, \mnras, 436, 1023

\bibitem[{{Becker} {et~al.}(2011){Becker}, {Bolton}, {Haehnelt}, \&
  {Sargent}}]{2011MNRAS.410.1096B}
{Becker}, G.~D., {Bolton}, J.~S., {Haehnelt}, M.~G., \& {Sargent}, W.~L.~W.
  2011, \mnras, 410, 1096

\bibitem[{{Becker} {et~al.}(2015){Becker}, {Bolton}, {Madau}, {Pettini},
  {Ryan-Weber}, \& {Venemans}}]{2015MNRAS.447.3402B}
{Becker}, G.~D., {Bolton}, J.~S., {Madau}, P., {et~al.} 2015, \mnras, 447, 3402

\bibitem[{{Becker} {et~al.}(2018){Becker}, {Davies}, {Furlanetto}, {Malkan},
  {Boera}, \& {Douglass}}]{2018arXiv180308932B}
{Becker}, G.~D., {Davies}, F.~B., {Furlanetto}, S.~R., {et~al.} 2018, ArXiv
  e-prints, arXiv:1803.08932

\bibitem[{{Bolton} \& {Haehnelt}(2007)}]{2007MNRAS.374..493B}
{Bolton}, J.~S., \& {Haehnelt}, M.~G. 2007, \mnras, 374, 493

\bibitem[{{Bolton} \& {Haehnelt}(2013)}]{2013MNRAS.429.1695B}
---. 2013, \mnras, 429, 1695

\bibitem[{{Bressan} {et~al.}(2012){Bressan}, {Marigo}, {Girardi}, {Salasnich},
  {Dal Cero}, {Rubele}, \& {Nanni}}]{2012MNRAS.427..127B}
{Bressan}, A., {Marigo}, P., {Girardi}, L., {et~al.} 2012, \mnras, 427, 127

\bibitem[{{Bruzual} \& {Charlot}(2003)}]{2003MNRAS.344.1000B}
{Bruzual}, G., \& {Charlot}, S. 2003, \mnras, 344, 1000

\bibitem[{{Caruana} {et~al.}(2014){Caruana}, {Bunker}, {Wilkins}, {Stanway},
  {Lorenzoni}, {Jarvis}, \& {Ebert}}]{2014MNRAS.443.2831C}
{Caruana}, J., {Bunker}, A.~J., {Wilkins}, S.~M., {et~al.} 2014, \mnras, 443,
  2831

\bibitem[{{Cen} {et~al.}(2009){Cen}, {McDonald}, {Trac}, \&
  {Loeb}}]{2009ApJ...706L.164C}
{Cen}, R., {McDonald}, P., {Trac}, H., \& {Loeb}, A. 2009, \apjl, 706, L164

\bibitem[{{Chabrier}(2003)}]{2003ApJ...586L.133C}
{Chabrier}, G. 2003, \apjl, 586, L133

\bibitem[{{Chardin} {et~al.}(2017){Chardin}, {Puchwein}, \&
  {Haehnelt}}]{2017MNRAS.465.3429C}
{Chardin}, J., {Puchwein}, E., \& {Haehnelt}, M.~G. 2017, \mnras, 465, 3429

\bibitem[{{Choi} {et~al.}(2017){Choi}, {Conroy}, \&
  {Byler}}]{2017ApJ...838..159C}
{Choi}, J., {Conroy}, C., \& {Byler}, N. 2017, \apj, 838, 159

\bibitem[{{Choi} {et~al.}(2016){Choi}, {Dotter}, {Conroy}, {Cantiello},
  {Paxton}, \& {Johnson}}]{2016ApJ...823..102C}
{Choi}, J., {Dotter}, A., {Conroy}, C., {et~al.} 2016, \apj, 823, 102

\bibitem[{{Choudhury} {et~al.}(2014){Choudhury}, {Puchwein}, {Haehnelt}, \&
  {Bolton}}]{2014arXiv1412.4790C}
{Choudhury}, T.~R., {Puchwein}, E., {Haehnelt}, M.~G., \& {Bolton}, J.~S. 2014,
  ArXiv e-prints, arXiv:1412.4790

\bibitem[{{Conroy} \& {Gunn}(2010)}]{2010ApJ...712..833C}
{Conroy}, C., \& {Gunn}, J.~E. 2010, \apj, 712, 833

\bibitem[{{Conroy} {et~al.}(2009){Conroy}, {Gunn}, \&
  {White}}]{2009ApJ...699..486C}
{Conroy}, C., {Gunn}, J.~E., \& {White}, M. 2009, \apj, 699, 486

\bibitem[{{D'Aloisio} {et~al.}(2018){D'Aloisio}, {McQuinn}, {Davies}, \&
  {Furlanetto}}]{2018MNRAS.473..560D}
{D'Aloisio}, A., {McQuinn}, M., {Davies}, F.~B., \& {Furlanetto}, S.~R. 2018,
  \mnras, 473, 560

\bibitem[{{D'Aloisio} {et~al.}(2015){D'Aloisio}, {McQuinn}, \&
  {Trac}}]{2015ApJ...813L..38D}
{D'Aloisio}, A., {McQuinn}, M., \& {Trac}, H. 2015, \apjl, 813, L38

\bibitem[{{D'Aloisio} {et~al.}(2017){D'Aloisio}, {Upton Sanderbeck}, {McQuinn},
  {Trac}, \& {Shapiro}}]{2017MNRAS.468.4691D}
{D'Aloisio}, A., {Upton Sanderbeck}, P.~R., {McQuinn}, M., {Trac}, H., \&
  {Shapiro}, P.~R. 2017, \mnras, 468, 4691

\bibitem[{{Davies} {et~al.}(2018{\natexlab{a}}){Davies}, {Becker}, \&
  {Furlanetto}}]{2017arXiv170808927D}
{Davies}, F.~B., {Becker}, G.~D., \& {Furlanetto}, S.~R. 2018{\natexlab{a}},
  \apj, 860, {155}

\bibitem[{{Davies} \& {Furlanetto}(2016)}]{2016MNRAS.460.1328D}
{Davies}, F.~B., \& {Furlanetto}, S.~R. 2016, \mnras, 460, 1328

\bibitem[{{Davies} {et~al.}(2016){Davies}, {Furlanetto}, \&
  {McQuinn}}]{2016MNRAS.457.3006D}
{Davies}, F.~B., {Furlanetto}, S.~R., \& {McQuinn}, M. 2016, \mnras, 457, 3006

\bibitem[{{Davies} {et~al.}(2018{\natexlab{b}}){Davies}, {Hennawi},
  {Ba{\~n}ados}, {Luki{\'c}}, {Decarli}, {Fan}, {Farina}, {Mazzucchelli},
  {Rix}, {Venemans}, {Walter}, {Wang}, \& {Yang}}]{2018arXiv180206066D}
{Davies}, F.~B., {Hennawi}, J.~F., {Ba{\~n}ados}, E., {et~al.}
  2018{\natexlab{b}}, ArXiv e-prints, arXiv:1802.06066

\bibitem[{{Deparis} {et~al.}(2018){Deparis}, {Aubert}, {Ocvirk}, {Chardin}, \&
  {Lewis}}]{2018arXiv180301634D}
{Deparis}, N., {Aubert}, D., {Ocvirk}, P., {Chardin}, J., \& {Lewis}, J. 2018,
  ArXiv e-prints, arXiv:1803.01634

\bibitem[{{Dotter}(2016)}]{2016ApJS..222....8D}
{Dotter}, A. 2016, \apjs, 222, 8

\bibitem[{{Doussot} {et~al.}(2017){Doussot}, {Trac}, \&
  {Cen}}]{2017arXiv171204464D}
{Doussot}, A., {Trac}, H., \& {Cen}, R. 2017, ArXiv e-prints, arXiv:1712.04464

\bibitem[{{Eldridge} \& {Stanway}(2009)}]{2009MNRAS.400.1019E}
{Eldridge}, J.~J., \& {Stanway}, E.~R. 2009, \mnras, 400, 1019

\bibitem[{{Faucher-Gigu{\`e}re} {et~al.}(2009){Faucher-Gigu{\`e}re}, {Lidz},
  {Zaldarriaga}, \& {Hernquist}}]{2009ApJ...703.1416F}
{Faucher-Gigu{\`e}re}, C.-A., {Lidz}, A., {Zaldarriaga}, M., \& {Hernquist}, L.
  2009, \apj, 703, 1416

\bibitem[{{Finlator} {et~al.}(2018){Finlator}, {Keating}, {Oppenheimer},
  {Dav{\'e}}, \& {Zackrisson}}]{2018arXiv180500099F}
{Finlator}, K., {Keating}, L., {Oppenheimer}, B.~D., {Dav{\'e}}, R., \&
  {Zackrisson}, E. 2018, ArXiv e-prints, arXiv:1805.00099

\bibitem[{{Furlanetto} \& {Oh}(2005)}]{2005MNRAS.363.1031F}
{Furlanetto}, S.~R., \& {Oh}, S.~P. 2005, \mnras, 363, 1031

\bibitem[{{Furlanetto} \& {Oh}(2009{\natexlab{a}})}]{furlanetto09}
---. 2009{\natexlab{a}}, \apj, 701, 94

\bibitem[{{Furlanetto} \& {Oh}(2009{\natexlab{b}})}]{2009ApJ...701...94F}
---. 2009{\natexlab{b}}, \apj, 701, 94

\bibitem[{{Furlanetto} \& {Stoever}(2010)}]{2010MNRAS.404.1869F}
{Furlanetto}, S.~R., \& {Stoever}, S.~J. 2010, \mnras, 404, 1869

\bibitem[{{Garzilli} {et~al.}(2012){Garzilli}, {Bolton}, {Kim}, {Leach}, \&
  {Viel}}]{2012MNRAS.424.1723G}
{Garzilli}, A., {Bolton}, J.~S., {Kim}, T.-S., {Leach}, S., \& {Viel}, M. 2012,
  \mnras, 424, 1723

\bibitem[{{George} {et~al.}(2015){George}, {Reichardt}, {Aird}, {Benson},
  {Bleem}, {Carlstrom}, {Chang}, {Cho}, {Crawford}, {Crites}, {de Haan},
  {Dobbs}, {Dudley}, {Halverson}, {Harrington}, {Holder}, {Holzapfel}, {Hou},
  {Hrubes}, {Keisler}, {Knox}, {Lee}, {Leitch}, {Lueker}, {Luong-Van},
  {McMahon}, {Mehl}, {Meyer}, {Millea}, {Mocanu}, {Mohr}, {Montroy}, {Padin},
  {Plagge}, {Pryke}, {Ruhl}, {Schaffer}, {Shaw}, {Shirokoff}, {Spieler},
  {Staniszewski}, {Stark}, {Story}, {van Engelen}, {Vanderlinde}, {Vieira},
  {Williamson}, \& {Zahn}}]{2015ApJ...799..177G}
{George}, E.~M., {Reichardt}, C.~L., {Aird}, K.~A., {et~al.} 2015, \apj, 799,
  177

\bibitem[{{Gnedin}(2016)}]{2016ApJ...833...66G}
{Gnedin}, N.~Y. 2016, \apj, 833, 66

\bibitem[{{G{\'o}rski} {et~al.}(2005){G{\'o}rski}, {Hivon}, {Banday},
  {Wandelt}, {Hansen}, {Reinecke}, \& {Bartelmann}}]{2005ApJ...622..759G}
{G{\'o}rski}, K.~M., {Hivon}, E., {Banday}, A.~J., {et~al.} 2005, \apj, 622,
  759

\bibitem[{{Gunawardhana} {et~al.}(2011){Gunawardhana}, {Hopkins}, {Sharp},
  {Brough}, {Taylor}, {Bland-Hawthorn}, {Maraston}, {Tuffs}, {Popescu},
  {Wijesinghe}, {Jones}, {Croom}, {Sadler}, {Wilkins}, {Driver}, {Liske},
  {Norberg}, {Baldry}, {Bamford}, {Loveday}, {Peacock}, {Robotham}, {Zucker},
  {Parker}, {Conselice}, {Cameron}, {Frenk}, {Hill}, {Kelvin}, {Kuijken},
  {Madore}, {Nichol}, {Parkinson}, {Pimbblet}, {Prescott}, {Sutherland},
  {Thomas}, \& {van Kampen}}]{2011MNRAS.415.1647G}
{Gunawardhana}, M.~L.~P., {Hopkins}, A.~M., {Sharp}, R.~G., {et~al.} 2011,
  \mnras, 415, 1647

\bibitem[{{Haardt} \& {Madau}(2012)}]{2012ApJ...746..125H}
{Haardt}, F., \& {Madau}, P. 2012, \apj, 746, 125

\bibitem[{{Hui} \& {Gnedin}(1997)}]{1997MNRAS.292...27H}
{Hui}, L., \& {Gnedin}, N.~Y. 1997, \mnras, 292, 27

\bibitem[{{Ir{\v s}i{\v c}} {et~al.}(2017){Ir{\v s}i{\v c}}, {Viel},
  {Haehnelt}, {Bolton}, {Cristiani}, {Becker}, {D'Odorico}, {Cupani}, {Kim},
  {Berg}, {L{\'o}pez}, {Ellison}, {Christensen}, {Denney}, \&
  {Worseck}}]{2017PhRvD..96b3522I}
{Ir{\v s}i{\v c}}, V., {Viel}, M., {Haehnelt}, M.~G., {et~al.} 2017, \prd, 96,
  023522

\bibitem[{{Keating} {et~al.}(2018){Keating}, {Puchwein}, \&
  {Haehnelt}}]{2018MNRAS.tmp..938K}
{Keating}, L.~C., {Puchwein}, E., \& {Haehnelt}, M.~G. 2018, \mnras,
  arXiv:1709.05351

\bibitem[{{Kroupa}(2001)}]{2001MNRAS.322..231K}
{Kroupa}, P. 2001, \mnras, 322, 231

\bibitem[{{Lidz} {et~al.}(2010){Lidz}, {Faucher-Gigu{\`e}re}, {Dall'Aglio},
  {McQuinn}, {Fechner}, {Zaldarriaga}, {Hernquist}, \&
  {Dutta}}]{2010ApJ...718..199L}
{Lidz}, A., {Faucher-Gigu{\`e}re}, C.-A., {Dall'Aglio}, A., {et~al.} 2010,
  \apj, 718, 199

\bibitem[{{Lidz} \& {Malloy}(2014)}]{2014ApJ...788..175L}
{Lidz}, A., \& {Malloy}, M. 2014, \apj, 788, 175

\bibitem[{{Lusso} {et~al.}(2015){Lusso}, {Worseck}, {Hennawi}, {Prochaska},
  {Vignali}, {Stern}, \& {O'Meara}}]{2015MNRAS.449.4204L}
{Lusso}, E., {Worseck}, G., {Hennawi}, J.~F., {et~al.} 2015, \mnras, 449, 4204

\bibitem[{{Ma} {et~al.}(2016){Ma}, {Hopkins}, {Faucher-Gigu{\`e}re}, {Zolman},
  {Muratov}, {Kere{\v s}}, \& {Quataert}}]{2016MNRAS.456.2140M}
{Ma}, X., {Hopkins}, P.~F., {Faucher-Gigu{\`e}re}, C.-A., {et~al.} 2016,
  \mnras, 456, 2140

\bibitem[{{Madau}(1995)}]{1995ApJ...441...18M}
{Madau}, P. 1995, \apj, 441, 18

\bibitem[{{Marks} {et~al.}(2012){Marks}, {Kroupa}, {Dabringhausen}, \&
  {Pawlowski}}]{2012MNRAS.422.2246M}
{Marks}, M., {Kroupa}, P., {Dabringhausen}, J., \& {Pawlowski}, M.~S. 2012,
  \mnras, 422, 2246

\bibitem[{{Mason} {et~al.}(2018){Mason}, {Treu}, {Dijkstra}, {Mesinger},
  {Trenti}, {Pentericci}, {de Barros}, \& {Vanzella}}]{2018ApJ...856....2M}
{Mason}, C.~A., {Treu}, T., {Dijkstra}, M., {et~al.} 2018, \apj, 856, 2

\bibitem[{{McGreer} {et~al.}(2015){McGreer}, {Mesinger}, \&
  {D'Odorico}}]{mcgreer15}
{McGreer}, I.~D., {Mesinger}, A., \& {D'Odorico}, V. 2015, \mnras, 447, 499

\bibitem[{{McQuinn}(2012)}]{2012MNRAS.426.1349M}
{McQuinn}, M. 2012, \mnras, 426, 1349

\bibitem[{{McQuinn}(2016)}]{2016ARA&A..54..313M}
---. 2016, \araa, 54, 313

\bibitem[{{McQuinn} {et~al.}(2007){McQuinn}, {Hernquist}, {Zaldarriaga}, \&
  {Dutta}}]{2007MNRAS.381...75M}
{McQuinn}, M., {Hernquist}, L., {Zaldarriaga}, M., \& {Dutta}, S. 2007, \mnras,
  381, 75

\bibitem[{{McQuinn} \& {Upton Sanderbeck}(2016)}]{2016MNRAS.456...47M}
{McQuinn}, M., \& {Upton Sanderbeck}, P.~R. 2016, \mnras, 456, 47

\bibitem[{{Mesinger} {et~al.}(2015){Mesinger}, {Aykutalp}, {Vanzella},
  {Pentericci}, {Ferrara}, \& {Dijkstra}}]{2015MNRAS.446..566M}
{Mesinger}, A., {Aykutalp}, A., {Vanzella}, E., {et~al.} 2015, \mnras, 446, 566

\bibitem[{{Miralda-Escud{\'e}} {et~al.}(2000){Miralda-Escud{\'e}}, {Haehnelt},
  \& {Rees}}]{2000ApJ...530....1M}
{Miralda-Escud{\'e}}, J., {Haehnelt}, M., \& {Rees}, M.~J. 2000, \apj, 530, 1

\bibitem[{{Miralda-Escud{\'e}} \& {Rees}(1994)}]{1994MNRAS.266..343M}
{Miralda-Escud{\'e}}, J., \& {Rees}, M.~J. 1994, \mnras, 266, 343

\bibitem[{{Mortlock} {et~al.}(2011){Mortlock}, {Warren}, {Venemans}, {Patel},
  {Hewett}, {McMahon}, {Simpson}, {Theuns}, {Gonz{\'a}les-Solares}, {Adamson},
  {Dye}, {Hambly}, {Hirst}, {Irwin}, {Kuiper}, {Lawrence}, \&
  {R{\"o}ttgering}}]{2011Natur.474..616M}
{Mortlock}, D.~J., {Warren}, S.~J., {Venemans}, B.~P., {et~al.} 2011, \nat,
  474, 616

\bibitem[{{Ouchi} {et~al.}(2010){Ouchi}, {Shimasaku}, {Furusawa}, {Saito},
  {Yoshida}, {Akiyama}, {Ono}, {Yamada}, {Ota}, {Kashikawa}, {Iye}, {Kodama},
  {Okamura}, {Simpson}, \& {Yoshida}}]{2010ApJ...723..869O}
{Ouchi}, M., {Shimasaku}, K., {Furusawa}, H., {et~al.} 2010, \apj, 723, 869

\bibitem[{{Paxton} {et~al.}(2011){Paxton}, {Bildsten}, {Dotter}, {Herwig},
  {Lesaffre}, \& {Timmes}}]{2011ApJS..192....3P}
{Paxton}, B., {Bildsten}, L., {Dotter}, A., {et~al.} 2011, \apjs, 192, 3

\bibitem[{{Paxton} {et~al.}(2013){Paxton}, {Cantiello}, {Arras}, {Bildsten},
  {Brown}, {Dotter}, {Mankovich}, {Montgomery}, {Stello}, {Timmes}, \&
  {Townsend}}]{2013ApJS..208....4P}
{Paxton}, B., {Cantiello}, M., {Arras}, P., {et~al.} 2013, \apjs, 208, 4

\bibitem[{{Paxton} {et~al.}(2015){Paxton}, {Marchant}, {Schwab}, {Bauer},
  {Bildsten}, {Cantiello}, {Dessart}, {Farmer}, {Hu}, {Langer}, {Townsend},
  {Townsley}, \& {Timmes}}]{2015ApJS..220...15P}
{Paxton}, B., {Marchant}, P., {Schwab}, J., {et~al.} 2015, \apjs, 220, 15

\bibitem[{{Pentericci} {et~al.}(2011){Pentericci}, {Fontana}, {Vanzella},
  {Castellano}, {Grazian}, {Dijkstra}, {Boutsia}, {Cristiani}, {Dickinson},
  {Giallongo}, {Giavalisco}, {Maiolino}, {Moorwood}, {Paris}, \&
  {Santini}}]{2011ApJ...743..132P}
{Pentericci}, L., {Fontana}, A., {Vanzella}, E., {et~al.} 2011, \apj, 743, 132

\bibitem[{{Planck Collaboration} {et~al.}(2016){Planck Collaboration}, {Adam},
  {Aghanim}, {Ashdown}, {Aumont}, {Baccigalupi}, {Ballardini}, {Banday},
  {Barreiro}, {Bartolo}, {Basak}, {Battye}, {Benabed}, {Bernard}, {Bersanelli},
  {Bielewicz}, {Bock}, {Bonaldi}, {Bonavera}, {Bond}, {Borrill}, {Bouchet},
  {Boulanger}, {Bucher}, {Burigana}, {Calabrese}, {Cardoso}, {Carron},
  {Chiang}, {Colombo}, {Combet}, {Comis}, {Couchot}, {Coulais}, {Crill},
  {Curto}, {Cuttaia}, {Davis}, {de Bernardis}, {de Rosa}, {de Zotti},
  {Delabrouille}, {Di Valentino}, {Dickinson}, {Diego}, {Dor{\'e}}, {Douspis},
  {Ducout}, {Dupac}, {Elsner}, {En{\ss}lin}, {Eriksen}, {Falgarone}, {Fantaye},
  {Finelli}, {Forastieri}, {Frailis}, {Fraisse}, {Franceschi}, {Frolov},
  {Galeotta}, {Galli}, {Ganga}, {G{\'e}nova-Santos}, {Gerbino}, {Ghosh},
  {Gonz{\'a}lez-Nuevo}, {G{\'o}rski}, {Gruppuso}, {Gudmundsson}, {Hansen},
  {Helou}, {Henrot-Versill{\'e}}, {Herranz}, {Hivon}, {Huang}, {Ili{\'c}},
  {Jaffe}, {Jones}, {Keih{\"a}nen}, {Keskitalo}, {Kisner}, {Knox},
  {Krachmalnicoff}, {Kunz}, {Kurki-Suonio}, {Lagache}, {L{\"a}hteenm{\"a}ki},
  {Lamarre}, {Langer}, {Lasenby}, {Lattanzi}, {Lawrence}, {Le Jeune},
  {Levrier}, {Lewis}, {Liguori}, {Lilje}, {L{\'o}pez-Caniego}, {Ma},
  {Mac{\'{\i}}as-P{\'e}rez}, {Maggio}, {Mangilli}, {Maris}, {Martin},
  {Mart{\'{\i}}nez-Gonz{\'a}lez}, {Matarrese}, {Mauri}, {McEwen}, {Meinhold},
  {Melchiorri}, {Mennella}, {Migliaccio}, {Miville-Desch{\^e}nes}, {Molinari},
  {Moneti}, {Montier}, {Morgante}, {Moss}, {Naselsky}, {Natoli}, {Oxborrow},
  {Pagano}, {Paoletti}, {Partridge}, {Patanchon}, {Patrizii}, {Perdereau},
  {Perotto}, {Pettorino}, {Piacentini}, {Plaszczynski}, {Polastri}, {Polenta},
  {Puget}, {Rachen}, {Racine}, {Reinecke}, {Remazeilles}, {Renzi}, {Rocha},
  {Rossetti}, {Roudier}, {Rubi{\~n}o-Mart{\'{\i}}n}, {Ruiz-Granados},
  {Salvati}, {Sandri}, {Savelainen}, {Scott}, {Sirri}, {Sunyaev}, {Suur-Uski},
  {Tauber}, {Tenti}, {Toffolatti}, {Tomasi}, {Tristram}, {Trombetti},
  {Valiviita}, {Van Tent}, {Vielva}, {Villa}, {Vittorio}, {Wandelt}, {Wehus},
  {White}, {Zacchei}, \& {Zonca}}]{2016A&amp;A...596A.108P}
{Planck Collaboration}, {Adam}, R., {Aghanim}, N., {et~al.} 2016, \aap, 596,
  A108

\bibitem[{{Price} {et~al.}(2016){Price}, {Trac}, \&
  {Cen}}]{2016arXiv160503970P}
{Price}, L.~C., {Trac}, H., \& {Cen}, R. 2016, ArXiv e-prints, arXiv:1605.03970

\bibitem[{{Puchwein} {et~al.}(2018){Puchwein}, {Haardt}, {Haehnelt}, \&
  {Madau}}]{2018arXiv180104931P}
{Puchwein}, E., {Haardt}, F., {Haehnelt}, M.~G., \& {Madau}, P. 2018, ArXiv
  e-prints, arXiv:1801.04931

\bibitem[{{Rapp} \& {Francis}(1962)}]{1962JChPh..37.2631R}
{Rapp}, D., \& {Francis}, W.~E. 1962, \jcp, 37, 2631

\bibitem[{{Salpeter}(1955)}]{1955ApJ...121..161S}
{Salpeter}, E.~E. 1955, \apj, 121, 161

\bibitem[{{Schneider} {et~al.}(2018){Schneider}, {Sana}, {Evans},
  {Bestenlehner}, {Castro}, {Fossati}, {Gr{\"a}fener}, {Langer},
  {Ram{\'{\i}}rez-Agudelo}, {Sab{\'{\i}}n-Sanjuli{\'a}n},
  {Sim{\'o}n-D{\'{\i}}az}, {Tramper}, {Crowther}, {de Koter}, {de Mink},
  {Dufton}, {Garcia}, {Gieles}, {H{\'e}nault-Brunet}, {Herrero}, {Izzard},
  {Kalari}, {Lennon}, {Ma{\'{\i}}z Apell{\'a}niz}, {Markova}, {Najarro},
  {Podsiadlowski}, {Puls}, {Taylor}, {van Loon}, {Vink}, \&
  {Norman}}]{2018Sci...359...69S}
{Schneider}, F.~R.~N., {Sana}, H., {Evans}, C.~J., {et~al.} 2018, Science, 359,
  69

\bibitem[{{Shapiro} \& {Giroux}(1987)}]{1987ApJ...321L.107S}
{Shapiro}, P.~R., \& {Giroux}, M.~L. 1987, \apjl, 321, L107

\bibitem[{{Shapiro} {et~al.}(2006){Shapiro}, {Iliev}, {Alvarez}, \&
  {Scannapieco}}]{2006ApJ...648..922S}
{Shapiro}, P.~R., {Iliev}, I.~T., {Alvarez}, M.~A., \& {Scannapieco}, E. 2006,
  \apj, 648, 922

\bibitem[{{Shapiro} {et~al.}(2004){Shapiro}, {Iliev}, \&
  {Raga}}]{2004MNRAS.348..753S}
{Shapiro}, P.~R., {Iliev}, I.~T., \& {Raga}, A.~C. 2004, \mnras, 348, 753

\bibitem[{{Springel}(2005)}]{2005MNRAS.364.1105S}
{Springel}, V. 2005, \mnras, 364, 1105

\bibitem[{{Stanway} {et~al.}(2016){Stanway}, {Eldridge}, \&
  {Becker}}]{2016MNRAS.456..485S}
{Stanway}, E.~R., {Eldridge}, J.~J., \& {Becker}, G.~D. 2016, \mnras, 456, 485

\bibitem[{{Stark} {et~al.}(2010){Stark}, {Ellis}, {Chiu}, {Ouchi}, \&
  {Bunker}}]{2010MNRAS.408.1628S}
{Stark}, D.~P., {Ellis}, R.~S., {Chiu}, K., {Ouchi}, M., \& {Bunker}, A. 2010,
  \mnras, 408, 1628

\bibitem[{{Taylor} \& {Lidz}(2014)}]{2014MNRAS.437.2542T}
{Taylor}, J., \& {Lidz}, A. 2014, \mnras, 437, 2542

\bibitem[{{Tittley} \& {Meiksin}(2007)}]{2007MNRAS.380.1369T}
{Tittley}, E.~R., \& {Meiksin}, A. 2007, \mnras, 380, 1369

\bibitem[{{Trac} \& {Cen}(2007)}]{2007ApJ...671....1T}
{Trac}, H., \& {Cen}, R. 2007, \apj, 671, 1

\bibitem[{{Trac} {et~al.}(2008){Trac}, {Cen}, \& {Loeb}}]{2008ApJ...689L..81T}
{Trac}, H., {Cen}, R., \& {Loeb}, A. 2008, \apjl, 689, L81

\bibitem[{{Trac} {et~al.}(2015){Trac}, {Cen}, \&
  {Mansfield}}]{2015ApJ...813...54T}
{Trac}, H., {Cen}, R., \& {Mansfield}, P. 2015, \apj, 813, 54

\bibitem[{{Trac} \& {Pen}(2004)}]{2004NewA....9..443T}
{Trac}, H., \& {Pen}, U.-L. 2004, \na, 9, 443

\bibitem[{{Upton Sanderbeck} {et~al.}(2016){Upton Sanderbeck}, {D'Aloisio}, \&
  {McQuinn}}]{2016MNRAS.460.1885U}
{Upton Sanderbeck}, P.~R., {D'Aloisio}, A., \& {McQuinn}, M.~J. 2016, \mnras,
  460, 1885

\bibitem[{{Venkatesan} \& {Benson}(2011)}]{2011MNRAS.417.2264V}
{Venkatesan}, A., \& {Benson}, A. 2011, \mnras, 417, 2264

\bibitem[{{White} {et~al.}(2003){White}, {Becker}, {Fan}, \&
  {Strauss}}]{2003AJ....126....1W}
{White}, R.~L., {Becker}, R.~H., {Fan}, X., \& {Strauss}, M.~A. 2003, \aj, 126,
  1

\bibitem[{{Zhang} {et~al.}(2018){Zhang}, {Romano}, {Ivison}, {Papadopoulos}, \&
  {Matteucci}}]{2018Natur.558..260Z}
{Zhang}, Z.-Y., {Romano}, D., {Ivison}, R.~J., {Papadopoulos}, P.~P., \&
  {Matteucci}, F. 2018, \nat, 558, 260

\end{thebibliography}

\begin{appendix}

\section{Numerical convergence and other tests}
\label{sec:convergence}

\subsection{Convergence}

\begin{figure}
\includegraphics[width=8.5cm]{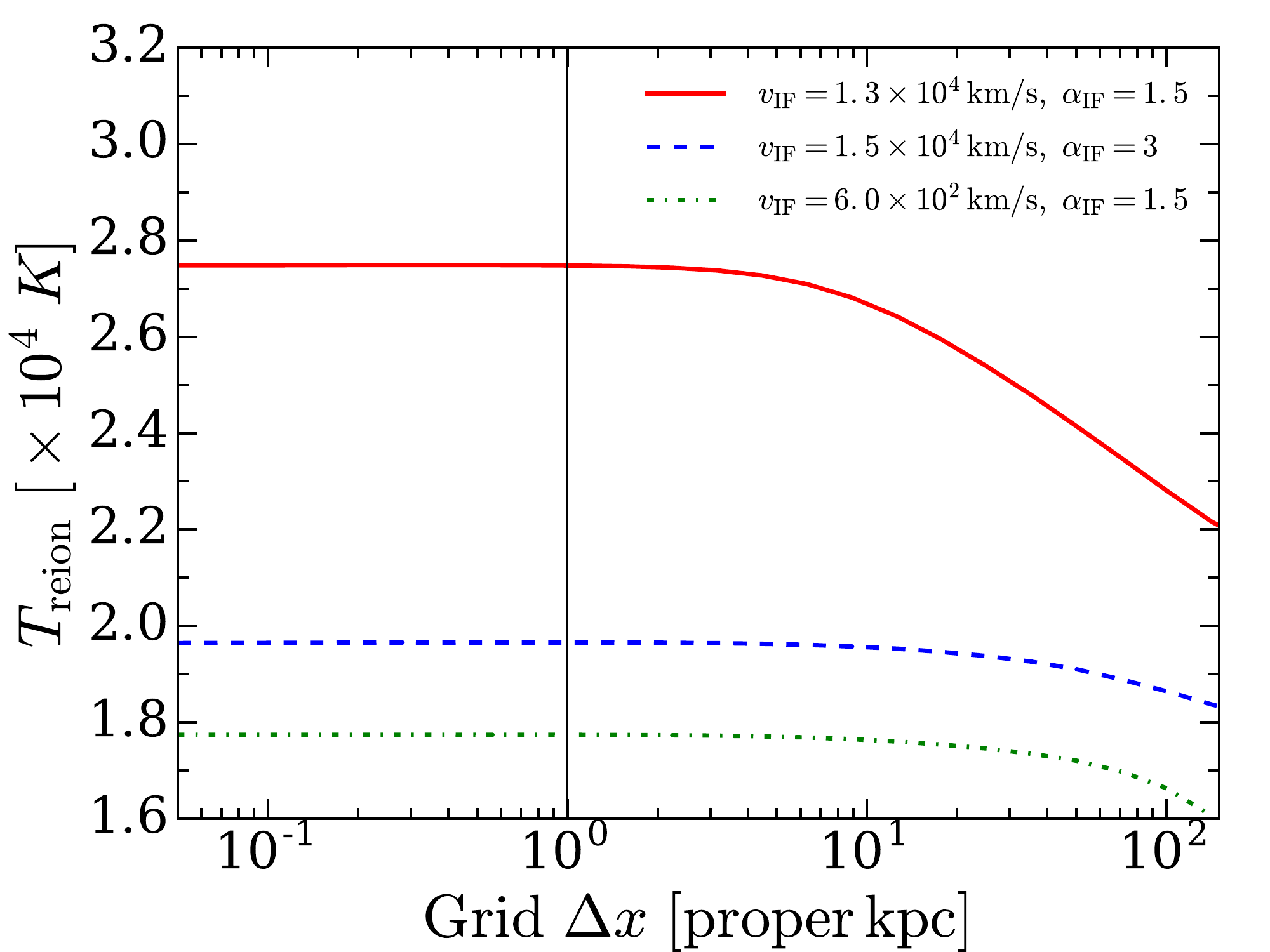}
\includegraphics[width=8.5cm]{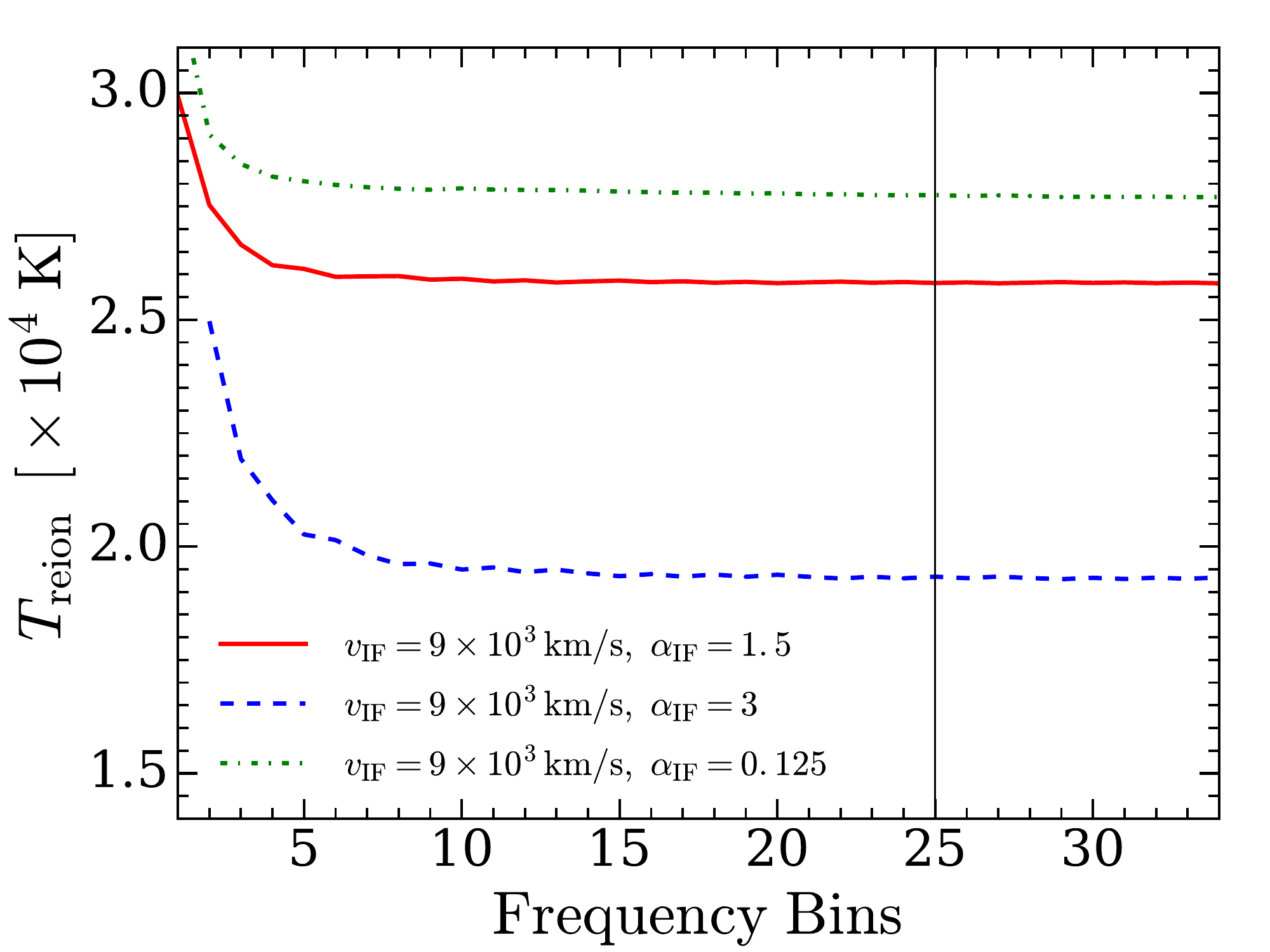}
\caption{Numerical convergence of our 1D RT simulations. Left panel: convergence with respect to the spatial grid cell size. The curves correspond to different I-front speeds and spectral indices as denoted in the legend.  The vertical line shows the cell size of $1$ proper kpc chosen for all of the runs in this paper. Right panel: convergence with respect to the number of RT frequency bins.  The bins are evenly spaced in log-space between 1 and 4 Ry. The vertical line corresponds to the number of frequency bins for all runs in this paper (25 bins).   }
\label{fig:convergence}
\end{figure}

Here we present numerical tests of our 1D RT simulation results.  We begin with numerical convergence.  The left panel of Fig. \ref{fig:convergence} demonstrates that our results are converged with respect to the spatial grid cell size, $\Delta x$.  The curves correspond to $\Treion$ at fixed $\vIF$ and $\spec$ over a range of simulation resolutions.  The vertical line shows the resolution adopted throughout the this paper, $\Delta x = 1$ proper kpc.  We note at the resolution requirements for $\Treion$ are most stringent for softer spectra because the I-fronts get thinner as $\spec$ increases.  Thus resolving the heating/cooling processes within the I-fronts requires finer spatial resolution as $\spec$ increases. Interestingly, the plot shows that $\Treion$ remains reasonably well-converged out to larger grid spacings of $\Delta x \approx 10$ proper kpc.  However, we caution against interpreting this convergence test in the context of other simulation codes, as the convergence properties likely differ considerably between different numerical approaches.   

In the right panel of Fig. \ref{fig:convergence}, we demonstrate the convergence of our results with respect to the number of frequency bins.  We divide up the spectrum into $N$ evenly spaced frequency bins in logarithmic space.  For all runs, the bins are bounded by 1 and 4 Ry, i.e. the low (high) end of the lowest (highest) frequency bin corresponds to 1 (4) Ry.  The vertical line in the plot corresponds to the fiducial value of 25 frequency bins chosen for all runs in this paper.     

\subsection{Test of the modified RT code}

\begin{figure}
\includegraphics[width=8.5cm]{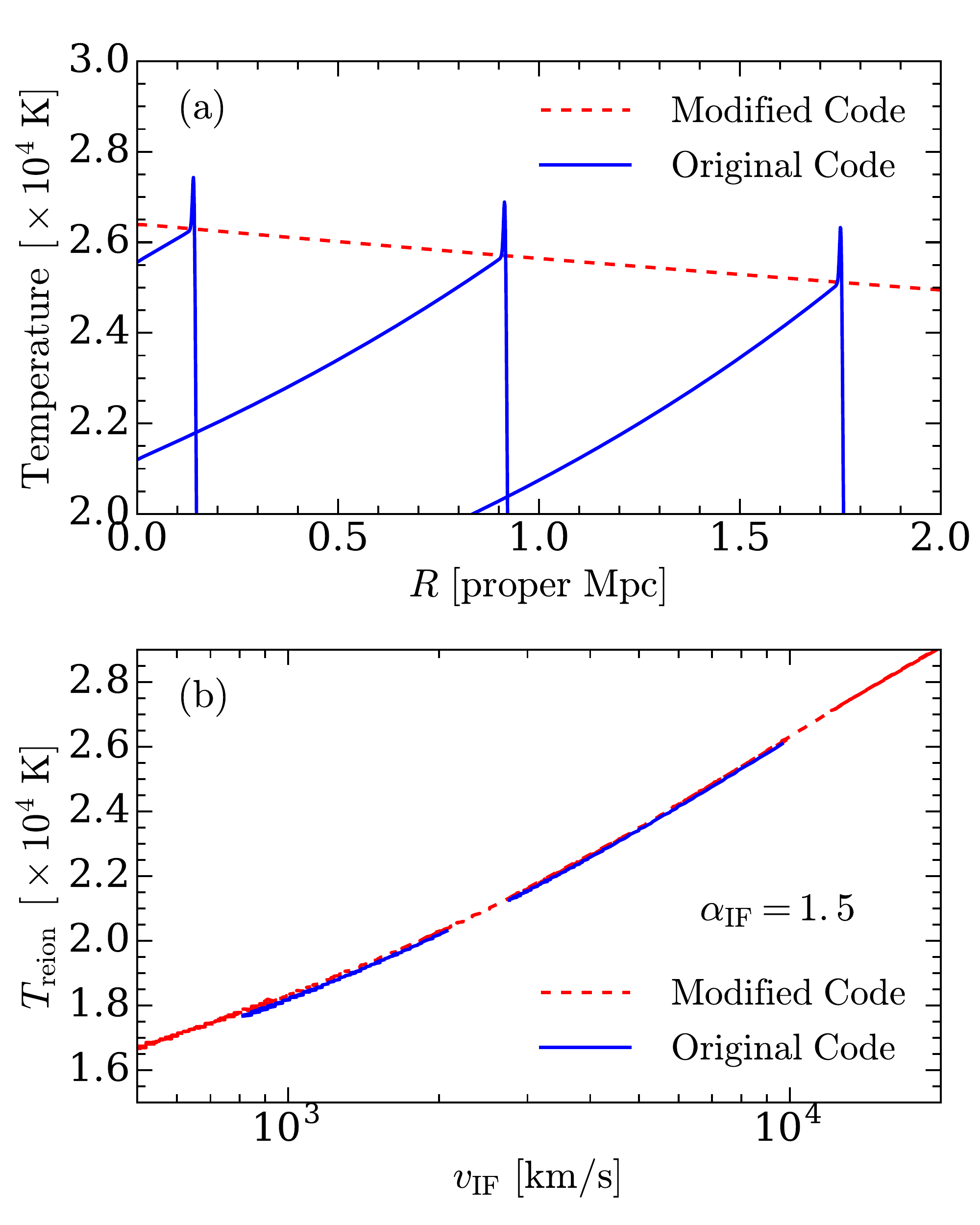}
\includegraphics[width=8.5cm]{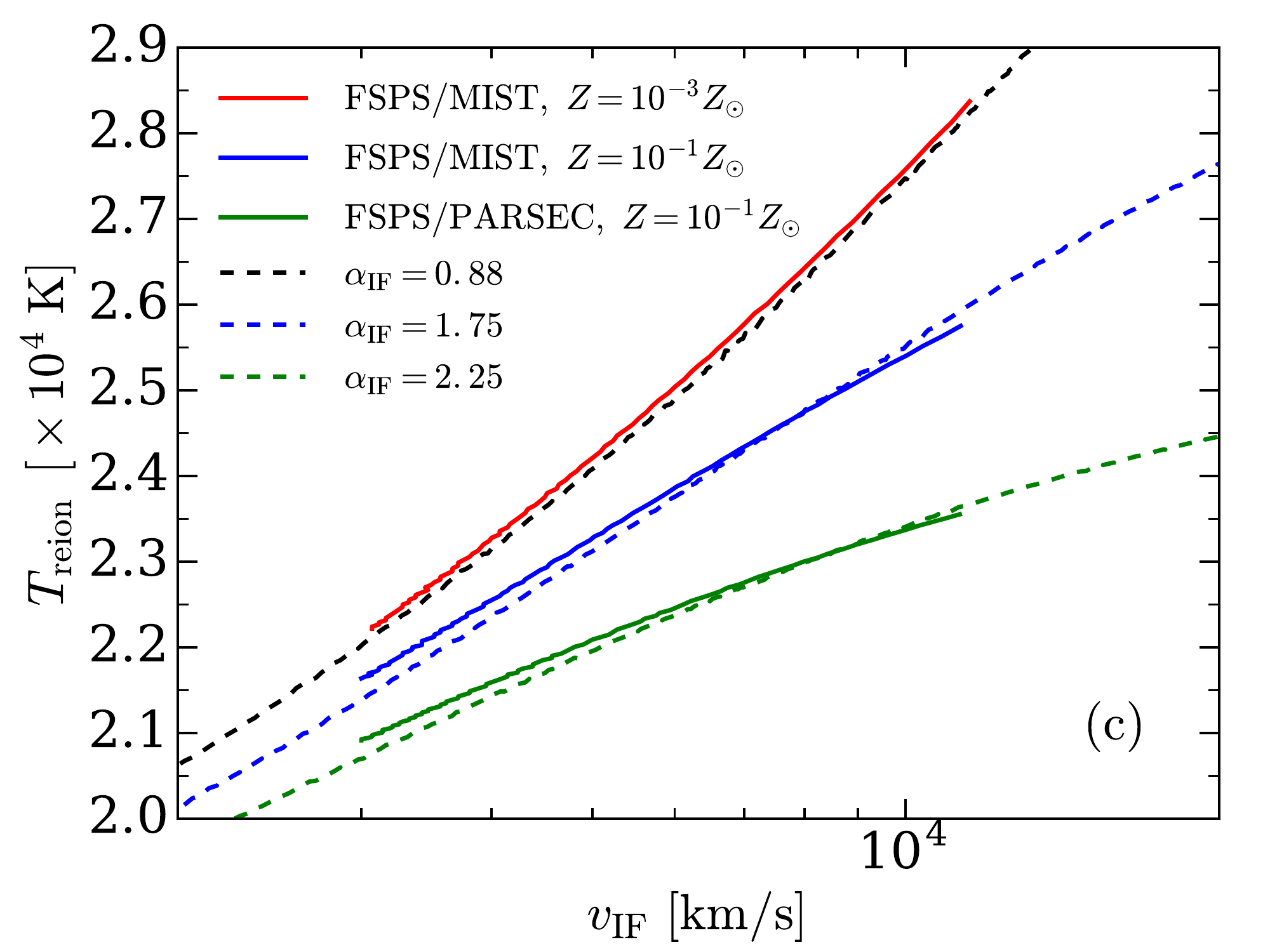}
\caption{ Tests of the 1D RT code. {\bf (a) } Visual comparison of $\Treion$ values obtained using the original and modified RT codes.  The solid curves show the gas temperature along the line of sight obtained from the original code at three snapshots in time: from left to right, $t=20$, 100, and 200 Myrs.  The dashed curve corresponds to the last snapshot of a simulation run with the modified code (but otherwise identical).  The modified code turns off the thermal evolution of the gas behind the front, allowing $\Treion$ to be obtained for all $R$ at the end of the simulation. {\bf (b) }  Quantitative comparison of $\Treion$ values.  Here we compare $\Treion$ vs. $\vIF$ for fixed $\spec=1.5$ using the original (solid) and modified (dashed) codes.    {\bf (c)} Post-I-front temperatures in RT runs with population synthesis model spectra. The solid curves correspond to the time-integrated FSPS spectra shown in Fig. \ref{FIG:spectra}.  The dashed curves correspond to the simple power-law models (with sharp cutoff at 4 Ry).}   
\label{fig:Treion_FSPS}
\end{figure}

As described in \S \ref{sec:methodology}, we modified the code of \citet{2016MNRAS.457.3006D} to make extracting $\Treion$ simpler, and to effectively define $\spec$ as the spectral index of the {\it incident} radiation.  Here we compare our code against the original, demonstrating that they produce nearly identical $\Treion$ values.  Let us begin by illustrating the utility of the modified code.  The solid curves in Figure \ref{fig:Treion_FSPS}a show the gas temperature in a simulation with the original code at three snapshots in time.  The sharp boundaries correspond to the locations of the I-front as it progresses from left to right.  The temperature peaks inside the I-front and the cooling behind it is driven primarily by Compton cooling and adiabatic expansion.   The dashed curve corresponds to the final snapshot of a simulation with the modified code, which turns off Compton and expansion cooling, as well as all thermal evolution behind the front (but is otherwise identical to the original simulation).  Note that the modified code effectively records the temperature immediately behind the I-front, i.e. the intersection of the dashed and solid curves, such that it can be simply read off from the final snapshot data.  In contrast, to obtain these temperatures from the original code, we would have to locate the I-front in a given snapshot and pick off the temperature behind the front, but before the cooling processes set in. It is difficult to obtain a robust prescription for doing this because the I-front widths vary significantly over the parameter space that we explore, and we must be careful to avoid the temperature structure within the I-front itself (i.e the peaks in Fig. \ref{fig:Treion_FSPS}a).  The modified code greatly simplifies this task.  In Fig. \ref{fig:Treion_FSPS}b, we compare $\Treion$ vs. $\vIF$ curves (for fixed $\spec=1.5$) obtained with the original and modified codes.  For the former, we record $\Treion$ by extracting the temperature 10 kpc behind the peak temperature. (Visually this provides a reasonable estimate for locating the back end of the I-front in this particular case.)  We note that the codes produce nearly identical results.        

\begin{figure}
\includegraphics[width=8.5cm]{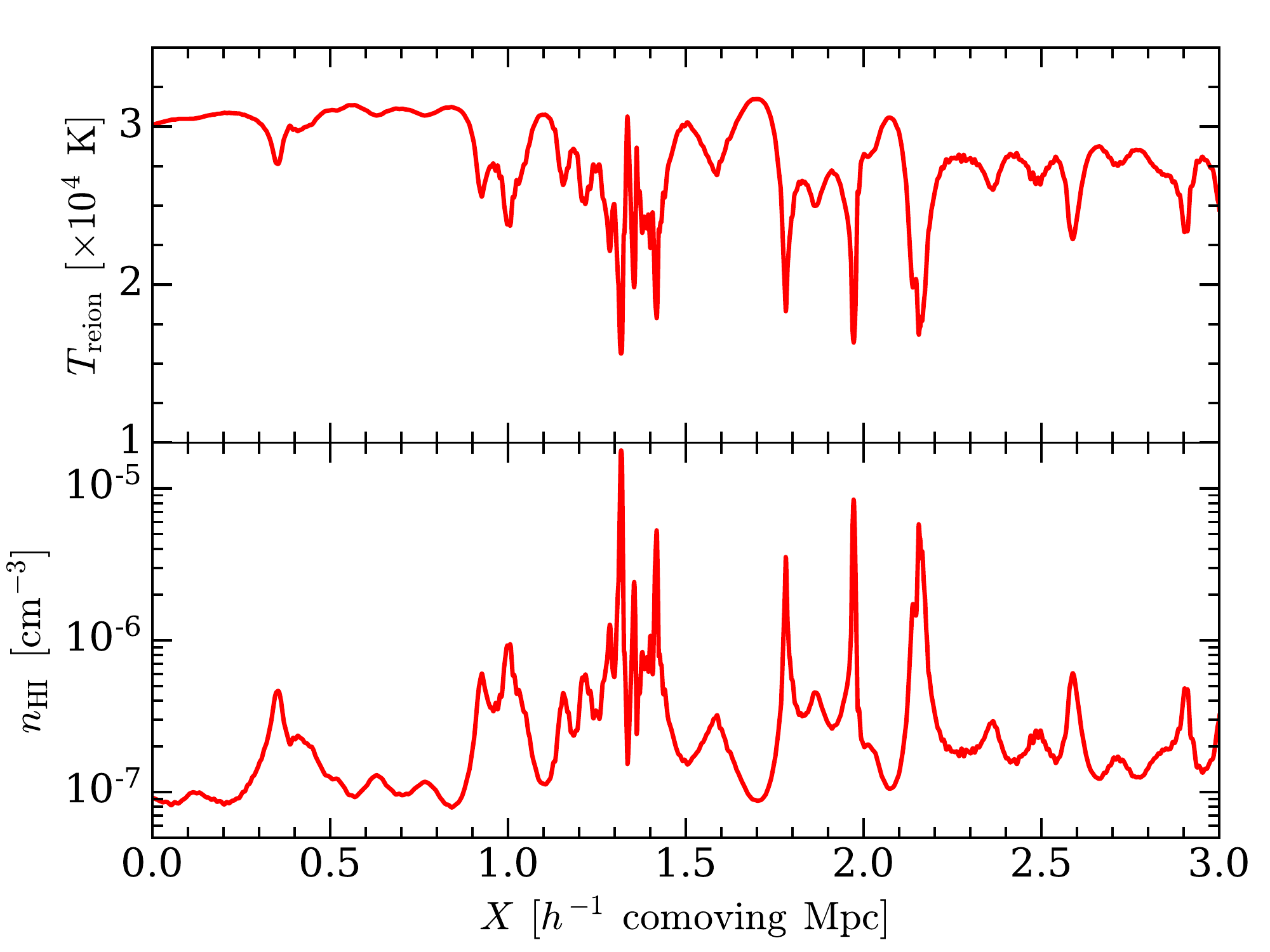}
\includegraphics[width=8.5cm]{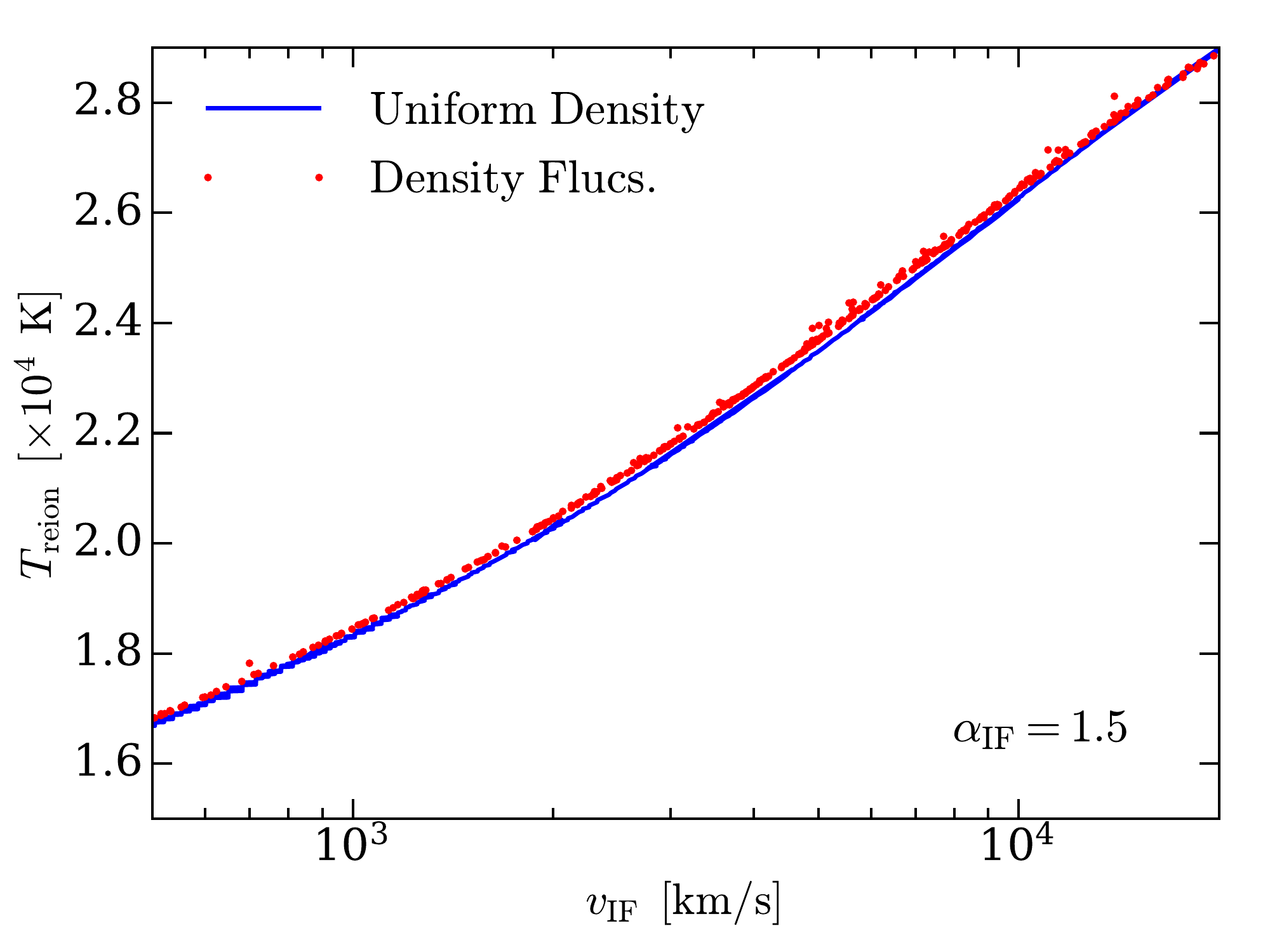}
\caption{ {Testing the effects of cosmological density fluctuations.  {\bf Left:} We performed RT simulations on 5 skewers extracted from a high-resolution cosmological simulation (see text for details).  Here we show example results for a segment of one skewer.  The top panel shows $\Treion$ along the skewer while the bottom panel shows the neutral hydrogen number density.  {\bf Right:} Although $\vIF$ depends on the densities encountered along the skewer, the $\Treion$ vs. $\vIF$ relation is unchanged by the presence of density fluctuations.  The red points show measurements along our 5 hydro skewers while the blue curve corresponds to our uniform-density runs.   } }
\label{fig:hydro_skewers}
\end{figure}

{

\subsection{Testing the effects of density fluctuations with cosmological simulations}
\label{sec:density_flucs}

All of the RT runs in this paper adopt a uniform IGM density.  In this section, we show by numerical tests that our main results are unaffected by the presence of cosmological density fluctuations.  We have performed a set of 5 RT runs on skewers through a high-resolution cosmological hydrodynamics simulation that was originally used in \citet{2016MNRAS.457.3006D}.  The simulation was run with the GADGET-3 code \citep{2005MNRAS.364.1105S} with a box size $L=3 h^{-1}$ Mpc and $N=2\times 512^3$ dark matter and gas particles.  To explore the maximum effect of density fluctuations, the simulation was run without photoheating from an ionizing background, but with a temperature floor of $T=50$ K.  We extracted 5 randomly drawn skewers at $z=7.2$ and performed RT in post-processing on them, adopting our fiducial source spectral index $\alpha = 1.5$.  

In contrast to our uniform density runs, the I-front velocities here can change quickly over short distances owing to the presence of density fluctuations.  Instead of measuring $\vIF$ in our usual way (which would require finite differencing over very short time scales), we use the flux method described in Appendix \ref{sec:fluxspeeds}, i.e. we obtain $\vIF$ from equation (\ref{EQ:fluxvel}).  The incident flux of ionizing photons, $F$, is measured at the rear of I-front where the neutral hydrogen fraction is $x_{\mathrm{HI}}=10$ \%.  We define the front boundary to be the location within the I-front where $x_{\mathrm{HI}}=50$ \%.  Thus, for the $n_{\mathrm{H}}$ that appears in equation (\ref{EQ:fluxvel}), we take the local neutral hydrogen density at this location and (after the front has passed) we measure $\Treion$ there as well.   In spite of our methods for eliminating the effects of spectral hardening by intervening gas between the source and the I-front (see \S \ref{sec:methodology}), some segments of the hydro sight lines are affected by optically thick over-densities that harden the incident spectral index, $\spec$.  This hardening results in somewhat higher $\Treion$ compared to our fiducial uniform density runs.  Since we would like to measure $\Treion$ for {\it fixed} $\spec$, we restrict our analysis here to segments for which $\spec$ has not been significantly hardened by optically thick absorbers.    

The left panel of Fig. \ref{fig:hydro_skewers} shows results for an example hydro skewer.  The top panel shows $\Treion$ for each cell along the skewer, while the bottom panel shows the neutral hydrogen density.   The I-front slows down (speeds up) in over(under-) dense regions, modulating $\Treion$ along the skewer.  Denser regions have lower $\Treion$.  { However, the right right panel of Fig. \ref{fig:hydro_skewers} shows that the mapping between $\vIF$ and $\Treion$ remains unaltered compared to our uniform-density runs.}   The red points in the right panel correspond to measurements of $(\vIF, \Treion)$ along our 5 hydro skewers, while the blue curve shows the corresponding result from our uniform density runs.  That the red points follow very tightly the uniform-density curve suggests that the contours of Fig. \ref{fig:Treion_contour} would be unaltered in the presence of cosmological density fluctuations.  This lack of sensitivity results from the fact that the relevant heating and cooling processes at a given location within the I-front depend only on the optical depth of the gas behind the location; they are independent of the structure of the intervening gas.  Additionally, \citet{2016MNRAS.457.3006D} argue that the I-fronts ``resolve" the density fluctuations such that the effective clumping factor for cooling processes is typically close to unity.  Because of the narrow width of the I-fronts in hydrogen column density, the density field never fluctuates appreciably on scales much smaller than the I-front.}


\subsection{RT runs with stellar population synthesis models}

Throughout this paper, we assume simple power-law spectra with sharp cutoffs at 4 Ry.  Conveniently, this allows us to parameterize the effects of the incident spectrum in terms of the spectral index, $\spec$.  But where do the integrated FSPS spectra in Fig. \ref{FIG:spectra} lie in this parameter space?  To address this question, we have also performed test runs using the FSPS spectra.   To make the results directly comparable to those of our power-law models, we have applied the same frequency binning to the FSPS spectra. Specifically, we re-bin the spectra into 25 logarithmically-spaced frequency bins between 1 and 4 Ry.  (We note that the FSPS spectra display a steep decline at 4 Ry such that photons above this energy contribute relatively little to the ionizing background in practice.)  The red and blue solid curves in Fig. \ref{fig:Treion_FSPS} show $\Treion$ versus $\vIF$ for the MIST models with $Z=10^{-3}Z_{\odot}$ and $Z=10^{-1}Z_{\odot}$, respectively.  The green solid curve corresponds to the PARSEC model with $Z=10^{-1}Z_{\odot}$.  The dashed curves show similar results assuming power-law spectra.  For the MIST models, the post-I-front temperatures are similar to those of power-law models with $\spec \approx 0.88$ and $1.75$ for $Z=10^{-3}Z_{\odot}$ and $Z=10^{-1}Z_{\odot}$, respectively.  For the PARSEC model, we find $\spec \approx 2.25$.  We note that these are similar to the effective spectral indices that are obtained by matching the mean energy per ionization in the optically thick limit (see discussion in \S \ref{sec:spec}). 

\section{Thermalization timescales within I-fronts}
\label{sec:thermalization}

All of our calculations assume that the timescale for photoelectrons to thermalize with the electrons, ions, and neutrals in an I-front is much shorter than the time over which the gas is within the front,
\begin{equation}
t_{\rm IF} = 9.8\times10^5 {\rm ~~ yr}\left(\frac{R_{\rm IF}}{\rm 10~ pkpc}\right) \left(\frac{v_{\rm IF}}{\rm 10^4~ km~s^{-1}}\right)^{-1}.
\label{EQ:tIF}
\end{equation}
Here, $R_{\rm IF}\sim 10$ (proper) kpc is a typical I-front width (motivated by Fig. \ref{fig:I-front_structure}), and $v_{\rm IF} \sim 10^4$km~s$^{-1}$ is a typical I-front speed in our cosmological RT simulations near the end of reionization (see \S \ref{sec:velocities}).  Following the discussion in the third paragraph of \S \ref{sec:Treion}, $t_{\rm IF} \propto \Delta^{-1}$ at fixed $v_{\rm IF}$. The rates in the ensuing paragraph also scale as $\Delta^{-1}$ and so we drop these dependences in our expressions for equilibration times.  In what follows, these times are defined as $[3/2 n_X k_b T] /|dU_{XY}/dt|$ where $n_X$ is the number density of the species of interest, $X$, and $dU_{XY}/dt$ is the energy exchange rate between species $X$ and $Y$, assuming $X$ is cold and $Y$ is at temperature T.  

We find that the assumption of a fast equilibration is justified.  Equilibration happens in the following sequence.  First, photoelectrons stream ahead of the I-front.  Their energy heats the electron bath in a short timescale, $t_{\rm photo-e, eq} \approx 50 ~(E/30\mathrm{eV})^{3/2} [x_i  Z_8^3]^{-1}$yr, where we have used the maximum energy of a photoelectron of $30$ eV (corresponding to the ionization of \HeI\ by a 4 Ry photon), and $Z_8 \equiv (1+ z)/8$.  (Little of the heat for such low energy photoelectrons goes {\it directly} into ionization or exciting atomic transitions, another assumption our calculations make.  We have tested this assumption in detail.)  Note that $t_{\rm photo-e, eq}$ should be evaluated at $x_i\sim 0.5$, which corresponds to the regime within an I-front where cooling becomes important \citep{2010MNRAS.404.1869F}.  The thermalized electrons then give their energy to the ions over again a relatively short timescale of $t_{\rm e-p, eq} \approx 3000~T_4^{3/2} [x_i  Z_8^3]^{-1}$yr, where $T_4 \equiv T/[10^4~{\rm K}]$.  Lastly, the neutrals are heated primarily by collisions with protons.  At the relevant temperatures, the collisional processes are dominated by resonant exchange of electrons (allowing e.g. hot ions to become hot neutrals as the kinetic energy is maintained), which we find is an order of magnitude more important than non-resonant collisions.  The equilibration timescale of the neutrals is $t_{nI} \approx 40,000 ~ T_4^{-1/2} [x_i  Z_8^3]^{-1}$yr \citep{1962JChPh..37.2631R, 1966P&amp;SS...14.1105B}, which is safely smaller than $t_{\rm IF}$ for all but the most relativistic speeds. 

\section{Estimating I-front speeds from photon fluxes}
\label{sec:fluxspeeds}

In this section, we describe an alternative method for estimating I-front speeds in cosmological RT simulations. Consider a plane-parallel I-front moving at speed $\vIF$ with respect to the frame of the gas.  The front is driven by impinging radiation with spectrum $S(\nu)$ and photon number flux, $F=\int_{\nu_{\mathrm{HI}}}^{\infty} d \nu S(\nu)$, where $\nu_{\mathrm{HI}}$ corresponds to the ionization threshold of hydrogen.  { Treating the I-front as a moving screen with velocity $\vIF$, the influx of neutral atoms on one side of the screen is balanced by the flux of ionizing photons on the other side}, $\mathcal{F} = (1-\vIF/c) \int_{\nu_{\mathrm{HI}}}^{\infty} \mathrm{d} \nu ~S(\nu)$.  Thus, the front velocity obeys {\citep{1994MNRAS.266..343M, 2006ApJ...648..922S}}

\begin{equation}
\mathcal{F} = n_{\mathrm{H}} (1+\chi) \vIF,
\label{EQ:fluxbalance}
\end{equation}
where $n_{\mathrm{H}}$ is the proper hydrogen number density, and the factor, $1+\chi$, accounts for the ionization of helium.  Here we assume that helium is singly ionized, in accordance with standard models of the reionization process, in which case $1+\chi = 1+n_{\mathrm{He}}/n_{\mathrm{H}} \approx 1.08$.  From equation (\ref{EQ:fluxbalance}), our estimator for the magnitude of the I-front velocity is {\citep{2006ApJ...648..922S}}

\begin{equation}
\vIF = \frac{c F }{F+c  n_{\mathrm{H}}(1+\chi)}.
\label{EQ:fluxvelB}
\end{equation}  
In \S \ref{sec:velocities}, we use equation (\ref{EQ:fluxvelB}) to calculate the distribution of $\vIF$ at a given epoch.   For a given $z$, we first use the $\zreion$ field to locate RT cells at the boundaries of the I-fronts, i.e. cells that are reionized between $z$ and $z+\delta z$.  In what follows, we use $\delta z = 0.02$, but we have tested that our results are insensitive to the exact choice.  Since equation (\ref{EQ:fluxvelB}) applies only in the limit of a sharp I-front boundary, we further select from the recently reionized cells those that have neutral hydrogen fractions $x_{\mathrm{HI}} < 0.01$ to avoid optically thick cells.  (Again, we have verified that our results are not sensitive to variations in the exact value of this threshold.)  We then compute the number flux $F$ in the selected cells using 

\begin{equation}
F = \frac{\Delta l_{\mathrm{rt}}}{\Delta t_{\mathrm{rt}}} \sum_{i=1}^{N_{\mathrm{freq}}} n_{\gamma,i}, 
\end{equation} 
where $n_{\gamma,i}$ is the photon number density in the $i$th frequency bin, $\Delta l_{\mathrm{rt}}$ is the proper RT cell length, and $\Delta t_{\mathrm{rt}}$ is the RT time step.  To obtain $n_{\mathrm{H}}$, we smooth the $n_{\mathrm{H}}$ field (which is stored at the hydro resolution) to the RT grid resolution by convolving with the coordinate-space top-hat function.  We have tested the accuracy of the above procedure against the results of a 1-dimensional version of the RadHydro code, in which the velocities of I-fronts can be directly measured. We find excellent agreement for all cases tested, including runs in which we vary the speed of light from $c_{\mathrm{sim}}/c=0.1$ to $1$.

\end{appendix}

\end{document}